\documentclass[aps,pre,twocolumn]{revtex4}

\usepackage{graphicx}
\expandafter\let\csname equation*\endcsname\relax
  \expandafter\let\csname endequation*\endcsname\relax
\usepackage{mathtools}
\usepackage{amsmath}
\usepackage{amssymb}
\usepackage{bm}
\usepackage{pifont}
\usepackage{color}
\usepackage[FIGTOPCAP]{subfigure}
\usepackage[latin1]{inputenc}

\begin{document}

\title[Tricritical points in a Vicsek model with bounded confidence]{
Tricritical points in a Vicsek model of self-propelled particles with bounded confidence}

\author{Maksym Romensky}
\affiliation{Department of Mathematics, Uppsala University, Box 480, Uppsala 75106, Sweden}
\affiliation{School of Physics, Complex and Adaptive Systems Lab, University College Dublin, Belfield, Dublin 4, Ireland}
\author{Vladimir Lobaskin}
\affiliation{School of Physics, Complex and Adaptive Systems Lab, University College Dublin, Belfield, Dublin 4, Ireland}
\author{Thomas Ihle}
\affiliation{Department of Physics, North Dakota State University, Fargo, ND 58108-6050, USA,}
\affiliation{Max-Planck-Institute for the Physics of Complex Systems, N{\"o}thnitzer Stra{\ss}e 38, 01187 Dresden, Germany}

\date{\today}
\begin{abstract}
We study the orientational ordering in systems of self-propelled particles with selective interactions. To introduce the
selectivity we augment the standard Vicsek model with a bounded-confidence collision rule: a given
particle only aligns to neighbors who have directions quite similar to its own. Neighbors whose directions deviate more than
a fixed restriction angle $\alpha$ are ignored. The collective dynamics of this systems is studied by agent-based simulations and
kinetic mean field theory. We demonstrate that the reduction of the
restriction angle leads to a critical noise amplitude decreasing monotonically with that angle, turning into a power law
with exponent $3/2$ for small angles. Moreover, for small system sizes we show that upon decreasing the restriction angle,
the kind of the transition to polar collective motion changes from continuous to discontinuous.
Thus, an apparent tricritical point with different scaling laws is identified and calculated analytically.
We investigate the shifting and vanishing of this point due to the formation of density bands as the system size is increased.
Agent-based simulations in small systems with large particle velocities show excellent agreement with the kinetic theory predictions.
We also find that at very small interaction angles the polar ordered phase becomes unstable with respect to the apolar phase.
We derive analytical expressions for the dependence of the threshold noise on the restriction angle.
We show that the mean-field kinetic theory also permits stationary nematic states below a restriction angle of $0.681\,\pi$.
We calculate the critical noise, at which the disordered state bifurcates to a nematic state, and find that it is always smaller than
the threshold noise for the transition from disorder to polar order. The disordered-nematic transition features two tricritical points:
At low and high restriction angle the transition is discontinuous but continuous at intermediate $\alpha$.
We generalize our results to systems that show fragmentation into more than two groups and obtain scaling laws for
the transition lines and the corresponding tricritical points. A novel numerical method to evaluate the nonlinear Fredholm integral
equation for the stationary distribution function is also presented. This method is shown to give excellent agreement with agent-based
simulations, even in strongly ordered systems at noise values close to zero.
\end{abstract}

%
\maketitle
%
%

\section{Introduction}

Dynamic self-organisation and, in particular, mechanisms of flocking behavior in groups of living species remain one of the
most intriguing problems at the interface of physics and biology. Numerous physical models of interacting self-propelled
agents have been proposed recently to study these phenomena (see review papers
\cite{toner.j:2005,ramaswami.s:2010,vicsek.t:2012}). The collective behavior much resembling the dynamics of living
organisms has also been observed in a variety of synthetic systems, generally referred to as active soft matter
\cite{paxton.w:2004,howse.j:2007,golestanian.r:2007,thakur.s:2011,taktikos.j:2012,theurkauff.i:2012,marchetti.c:2013}.
The level of consciousness of the individuals apparently plays minor role for the large scale dynamics as the same general
principles that apply to groups of animals or cells seem to govern also human social phenomena, traffic, robotics, and
decision making \cite{helbing.d:2000,helbing.d:2001,deffuant.g:2001,weidlich.w:2002}. Therefore, similar modelling
approaches are employed to describe their generic dynamic properties.

One of the simplest and earliest models to describe the collective motion of self-propelled agents has been introduced by
Vicsek in 1995 \cite{vicsek.t:1995}. In this original paper, a second order phase transition between the orientationally
ordered, globally aligned motion state and disordered state was claimed. The continuous nature of the transition between
these states was also supported by a number of publications from the same group involving original Vicsek model (VM) with angular
noise \cite{czirok.a:1997,czirok.a:2000}. However, the nature of the transition has been questioned in a number of later
studies \cite{gregoire:2004,chate.h:2008}. Chat\'{e} et al. have demonstrated \cite{chate.h:2008} that there exists a
critical system size $L^*$, beyond which a discontinuous, or first order transition is observed. The dependence of this
critical system size on particle density has been theoretically estimated in Ref. \cite{ihle.t:2011}.
It has also been reported that the kind of the transition seems to depend on particle velocities \cite{baglietto.g:2009} and on the way,
in which the noise is introduced into the system \cite{aldana.m:2007}. All above mentioned factors can result in a behavior,
which can be associated with instabilities leading to co-existence of the ordered and disordered phases at the transition point,
or, in other words, to a first order phase transition \cite{nagy.m:2007}. The discontinuous character of the transition has
also been elucidated from analysis of Ising-like 1D models of flocking \cite{solon.ap:2013}.

The contribution of range and symmetry of aligning interactions to the instabilities remains relatively poorly understood.
Recent numerical studies demonstrated essential differences between the symmetry of the ordered phase and stability of density
bands depending on whether the active agents have polar or apolar alignment mechanism \cite{peruani.f:2011}.
In this paper we study the role of selectivity of interactions in the ordering of the Vicsek model (VM).
This property has a number of important applications.
For example, the selectivity of the interaction
has previously been introduced in the study of the social models such as the bounded confidence model \cite{deffuant.g:2001}.
In a situation where the agents are prepared to align themselves only with the fellow individuals with the opinion vector not
too different from their own one, the restrictive rules can become crucial for the polarization of the group and the
collective decision making. The bounded confidence rules can be introduced into the Vicsek model as a restriction on the
angle of interaction, which makes the alignment of the individual with too different directions of motion impossible
\cite{deffuant.g:2001,ben-naim.e:2003}.
Given that the Vicsek model was originally introduced to model the behavior of social agents such as birds and fish,
such a selectivity rule could make their description more realistic.

Another motivation to study an interaction rule that is sensitive to orientation differences are the very recent experiments
by Lu {\em et al.} \cite{lu.s:2013} on the collective behavior of {\em Bacillus subtilis} in the presence of a photosensitizer.
To account for the cell-to-cell interaction via intercellular flagella bundling, the authors propose a Vicsek-like model
where the interaction among neighbor agents becomes weaker with increasing orientation difference.
Since our model explores an extreme version of this interaction,
the kinetic theory from this paper might become useful for a better analytical understanding of these experiments.

In this paper, we derive the ordering behavior in the Vicsek model with bounded confidence from the microscopic kinetic
equations and demonstrate how a qualitatively new behavior arises from the interaction selectivity.
In particular, we find within homogeneous mean-field theory that at strong selectivity,
the transition from a disordered state to a state of polar order is discontinuous.
If one increases the restriction angle $\alpha$
beyond
a so-called tricritical point the transition becomes continuous.
We also analyze nematic phases and fragmented states which consist of several aligned groups
moving in different directions.
A scaling law for the transition from disorder to an ordered state with $p=2,3,\ldots$
fragments is calculated and two different tricritical points
are identified.
Furthermore, we present a novel numerical method to accurately evaluate the nonlinear Fredholm integral equation for stationary distribution
functions.
The paper is
organized as follows: in Section \ref{theory} we introduce the kinetic theory of the Vicsek model and predict its phase
behavior, describe the numerical solution methodology for the kinetic equations and the settings of the agent-based simulation.
In Section \ref{results}, we present the numerical and analytical results for the model: phase diagrams and data on the
ordering behavior. Finally, in Sections \ref{discussion} and \ref{summary} we discuss the significance of main findings of
the work and summarize the results.
Details about the calculations of coupling integrals are relegated to Appendix A.
In Appendix B we provide a general discussion of terms such as ``phase transition'' that are borrowed from
equilibrium statistical mechanics but here are applied to finite nonequilibrium systems.

\section{Theory and simulation settings}
\label{theory}
\subsection{Model}

\begin{figure}
\centering
\includegraphics[width=5.0cm,clip]{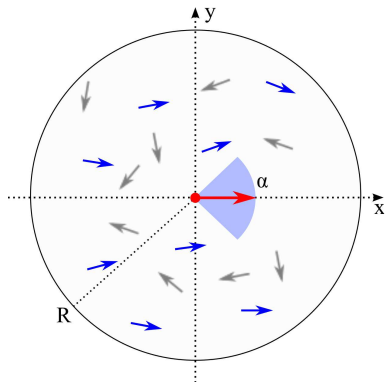}
\caption{The interaction parameters in the restricted angle Vicsek model. The particle aligns itself only with those neighbors (blue arrows) whose relative angle of motion is less or equal to $\alpha$.}
\label{fig:vicsek_ra_intparam}
\end{figure}

We consider a two-dimensional model with $N$ point particles at number density $\rho$, which move at constant speed $v_0$.
The particles with positions ${\bf x}_i(t)$ and velocities ${\bf v}_i(t)$ undergo discrete-time dynamics
with time step $\tau$. The evolution consists of two steps: streaming and collision.
In the streaming step all positions are updated according to
\begin{equation}
\label{VM_UPDATE}
{\bf x}_i(t+\tau)={\bf x}_i(t)+\tau {\bf v}_i(t)\,.
\end{equation}
In the subsequent collision step,
the directions $\theta_i$ of
the velocity vectors change.
Similar to the standard VM,
particles align with their neighbors within a fixed distance $R$. However, the interaction in the present paper is selective
such that the particles align only with those neighbors whose direction of motion deviates by an angle less than some fixed value $\alpha$ from their own velocity vector (see Fig. \ref{fig:vicsek_ra_intparam}).
In this implementation, the Vicsek model becomes similar to so-called ``Bounded confidence'' models, commonly used in social sciences to study opinion dynamics \cite{deffuant.g:2001,ben-naim.e:2003}.
It simulates the common social tendency to disregard opinions that appear too extreme with regard to their own perspective.
Thus, the parameter $\alpha$ can be interpreted as the degree of ignorance of a population of self-propelled agents.
In particular,
a circle of radius $R$ is drawn around a given particle and the average angle $\Phi_i$ of motion
of the
particles
within the circle is determined
according to
\begin{equation}
\Phi_i={\rm arctan} \left [\sum_{\{j\}} {\rm sin}(\theta_j)/\sum_{\{j\}} {\rm cos}(\theta_j) \right ]\,,
\end{equation}
where particles $j$ whose inner product
${\bf v}_j\cdot {\bf v}_i$ is smaller than $v_0^2\cos{\alpha}$ are excluded from the summation.
In an extreme case, it is possible that even if particle $i$ has many neighbors in its collision circle,
all are rejected due to too large differences, and thus $\Phi_i=\theta_i$.
The regular VM is recovered in the limit $\alpha=\pi$.
Once the average angles $\Phi_i$ are known, the new particle directions
follow as
\begin{equation}
\theta_i(t+\tau)=\Phi_i+\xi_i
\end{equation}
 where $\xi_i$ is a random number which is
uniformly distributed in
the interval $[-\eta/2,\eta/2]$.
Note, that the updated positions ${\bf x}_i(t+\tau)$ (and not the old locations ${\bf x}_i(t)$)
are used to determine the average angles of motion $\Phi_i$.
This corresponds to the so-called forward updating rule of the standard VM, as defined in Refs. \cite{huepe.c:2008,baglietto.g:2009}).

Note, that our model is qualitatively different from
models with a finite sensing region such as the ``blind-spot'' models introduced
by Couzin {\em et al.} \cite{couzin.id:2002}, and used, for example, in Refs. \cite{lukeman.r:2010,wood.aj:2007}.
In contrast to these models, particle exclusion is based on
velocity differences and not spatial location.
It can be easily shown that Vicsek-like models with rear blind sectors do not show tricritical points
in homogeneous mean-field theory.
However, the version of the VM presented here is related to the one introduced by Lu {\em et al.} \cite{lu.s:2013},
where instead of completely excluding misaligned partners, the interactions are weighted by the orientation difference.

\subsection{Kinetic theory}
\label{sec:kinetic_theory}
\subsubsection{Deriving the kinetic equation}
Recently, a kinetic theory formalism for self-propelled particles
has been introduced that can handle discrete time dynamics
and ``exotic'' collision rules such as genuine multi-body and
topological interactions
\cite{ihle.t:2011,chou.yl:2012,ihle.t.1:2014}.
It has been shown \cite{ihle.t:2013} that this approach is able to quantitatively reproduce
the results of agent-based simulations, in the limit of large mean free path, $\lambda=v_0\tau$, compared to the radius of interaction $R$.
In this chapter we will adapt this method to the restricted-angle model.
The formalism starts with an evolution equation for a Markov chain in phase space,
\begin{equation}
\label{MASTER1}
P({\bf B},t+\tau)=\int P({\bf A},t)\;w_{AB}\; d{\bf A}\,.
\end{equation}
where $P$ is the $N$-particle probability density referring to an ensemble of independent Vicsek systems, which are initialized at time $t=0$ with some initial probability density $P_0$. This initial density is assumed to be symmetric against permuting particle indices. Equation \eqref{MASTER1} describes the transition from a microscopic state ${\bf A}$
to the state ${\bf B}$ during one time step with transition probability $w_{AB}$.
The microscopic state of the system
at time $t+\tau$ is given by
the $3N$-dimensional
vector,
$B\equiv(\theta^{(N)}, {\bf X}^{(N)})$, where
$\theta^{(N)}\equiv(\theta_1,\theta_2,\ldots, \theta_N)$ contains the directions of motion of all $N$ particles, and
${\bf X}^{(N)}\equiv({\bf x}_1,{\bf x}_2,\ldots, {\bf x}_N)$ describes all particle positions.
The initial microscopic state at time $t$ is denoted as
${\bf A}\equiv(\tilde{\theta}^{(N)}, {\bf \tilde{X}}^{(N)})$.
The integral over the initial state translates to
$\int\,d{\bf A}\equiv \prod_{i=1}^N\int_0^{2\pi}\,d\tilde{\theta}_i\int\,d{\bf \tilde{x}}_i$.
Pre-collisional angles and positions are given by $\tilde{\theta}_j$ and ${\bf \tilde{x}}_i$, respectively.
The transition probability $w_{AB}$ contains the microscopic collision rules,
\begin{eqnarray}
\nonumber
w_{AB}&=&
{1\over \eta^N}
\prod_{i=1}^N
\delta({\bf \tilde{x}}_i-[{\bf x}_i-\tau {\bf v}_i]) \\
\label{TRANS1}
& &\times \int_{-\eta/2}^{\eta/2}
d\xi_i\,
\hat{\delta}(\theta_i-\xi_i-\Phi_i)\,.
\label{TRANSIT1}
\end{eqnarray}
Here,
$\hat{\delta}(x)=\sum_{m=-\infty}^{\infty}\delta(x+2\pi m)$,
is the periodically continued delta function,
which accounts for the $2\pi$-periodicity of angles.
The particle velocities
${\bf V}^{(N)}\equiv({\bf v}_1,{\bf v}_2,..., {\bf v}_N)$, are given in terms
of angle variables, ${\bf v}_i=v_0(\cos{\theta_i}, \sin{\theta_i})$.

The kinetic equation for the $N$-particle probability density, Eq. (\ref{MASTER1}) is
exact but intractable without simplification.
Here, as done, for example, in \cite{ihle.t:2011},
we use Boltzmann's molecular chaos approximation
by assuming that the particles are uncorrelated prior to a collision,
which amounts to a factorization of the $N$-particle probability into a product of one-particle probabilities,
$P(\theta^{(N)}, {\bf X}^{(N)})
=\prod_{i=1}^N P_1(\theta_i, {\bf x}_i)$.
This approximation
can be justified at moderate and large noise strength $\eta$ and
when the mean free path (mfp) is large compared to the radius of interaction $R$.
Here, the mfp is defined as the distance a particle travels between collisions,
$\tau\,v_0$, and is density-independent due to the discrete nature of the dynamics.
More details on the validity of molecular chaos and a general discussion of kinetic theory approaches to Vicsek-style models
can be found in Refs. \cite{ihle.t.1:2014,ihle.t.2:2014,peshkov.a:2014,chou.yl:2014}.

Because molecular chaos neglects pre-collisional correlations, it leads to a mean-field theory.
To derive this mean-field theory, we follow Refs. \cite{ihle.t:2011,malevanets.a:1999,ihle.t:2009} and multiply Eq. (\ref{MASTER1}) by
the phase space density $\sum_i\delta({\bf v}-{\bf v}_i)\delta({\bf x}-{\bf x}_i)$.
A subsequent integration
over all particle positions ${\bf x}_i$ and angles $\theta_i$ leads,
in the large $N$ limit,
to an Enskog-like kinetic equation for
the one-particle distribution function,
$f(\theta,{\bf x},t)=NP_1(\theta,{\bf x},t)$,
\begin{eqnarray}
\nonumber
& &f(\theta, {\bf x}+\tau {\bf v},t+\tau)=
{1\over \eta}
\int_{-\eta/2}^{\eta/2}
d\xi
\bigg\langle \bigg\langle
\sum_{n=1}^N
{{\rm e}^{-M_R}\over (n-1)!}
\\
\label{ENSKOG1}
& &\times f(\tilde{\theta}_1,{\bf x}, t)
\,\hat{\delta}(\theta-\xi-\Phi_1)
\,\prod_{i=2}^n f( \tilde{\theta}_i, {\bf x}_i,t)
\bigg\rangle_{\tilde{\theta}} \bigg\rangle_x
\end{eqnarray}
where
$M_R({\bf x},t)=\int_R\rho({\bf y},t)\,d{\bf y}$
is the average number of particles in a circle of radius $R$ centered around ${\bf x}$ and
can be position-dependent.
The subscript ``$R$'' at the integral denotes integration over this circle.
The local particle density $\rho$ is given as a moment of the distribution function,
$\rho({\bf x},t)=\int_0^{2\pi} f(\theta,{\bf x},t)\,d\theta$;
$\langle ... \rangle_x=\int_R... \,d{\bf x}_2\,d{\bf x}_3...d{\bf x}_n$
denotes the integration over all positions, $n-1$ particles can assume within the interaction circle;
$\langle ... \rangle_{\tilde{\theta}}=\int_0^{2\pi} ... d \tilde{\theta}_1 d \tilde{\theta}_2 ...
d \tilde{\theta}_n $
is the average over all pre-collisional angles
of $n$ particles in the interaction circle.
In Eq. (\ref{ENSKOG1}) particle $1$ is assumed to be the focal particle.
It is fixed at position ${\bf x}$ and particles $2,3\ldots n$ are supposed to be its neighbors.
More details on how to interpret equations similar to Eq. (\ref{ENSKOG1}), can be found in Ref. \cite{ihle.t.1:2014}.

\subsubsection{Solving the kinetic equation}

We restrict ourselves to spatially homogeneous solutions of the Enskog-like equation.
Then Eq. (\ref{ENSKOG1})
becomes
\begin{equation}
\label{FRED1}
f(\theta,t+\tau)=I[f(\theta,t)]
\end{equation}
with the simpler collision integral
\begin{eqnarray}
\nonumber
& &I[f(\theta,t)]=
{1\over \eta}
\int_{-\eta/2}^{\eta/2}
d\xi\,
\sum_{n=1}^N\,
{S^{n-1} {\rm e}^{-M}\over (n-1)!}
\\
\label{FRED2}
&\times & \bigg ( \prod_{i=1}^n f(\tilde{\theta}_i,t)\,d\tilde{\theta}_i\bigg )
\,\hat{\delta}[\theta-\xi-\Phi_1(\tilde{\theta}_1,\ldots\tilde{\theta}_n)]
\end{eqnarray}
where $S=\pi R^2$ is the area of the collision circle and the average particle number in this circle, $M=S\rho_0$, is proportional to
the constant particle number density $\rho_0$.
For stationary solutions where $f(\theta,t+\tau)=f(\theta,t)$, Eq. (\ref{FRED1}) constitutes
a nonlinear Fredholm integral equation of the second kind with a singular kernel.
Further below, we will present a novel numerical method to solve this equation with high precision.
In principle, spatially inhomogeneous solutions, including steep soliton-like waves,
could be obtained by the Lattice-Boltzmann-like method of
Ref. \cite{ihle.t:2013} but this is beyond the scope of this paper. Instead, we will handle inhomogeneous states by
agent-based simulations only.

A convenient starting point for analytical and numerical studies of Eq. (\ref{FRED1}) is the angular Fourier expansion
of both the distribution function and the collision integral,
\begin{eqnarray}
\nonumber
f&=&\sum_{k=0}^\infty g_k(t)\,{\rm cos}(k\theta) \\
\label{FOURIER1}
I[f]&=&\sum_{k=0}^\infty C_k(t)\,{\rm cos}(k\theta)\,.
\end{eqnarray}
The emergence of a globally ordered state breaks the rotational symmetry. Particles start to
flow preferentially in a common but arbitrary direction $\theta_0$.
Since we are only interested in steady states of homogeneous systems in the thermodynamic limit, it suffices
to set $\theta_0=0$ and to
only keep the cosine terms in the Fourier expansion \cite{FOOT2.ihle:2014}.
The zeroth mode $g_0$ is proportional to the average density,
\begin{equation}
\label{G0_DEF}
g_0={\rho_0 \over 2 \pi}
\end{equation}
because
$\int f(\theta)\,d\theta=\rho_0$. The first mode, $g_1$,
serves as polar order parameter and is determined by the average $x$-component of the momentum density ${\bf w}$,
\begin{eqnarray}
\nonumber
w_x&=&\int_0^{2\pi}v_x(\theta)f(\theta)\,d\theta \\
\label{G1_DEF}
g_1&=&{w_x\over v_0 \pi}
\end{eqnarray}
with $(v_x,v_y)=v_0(\cos{\theta},\sin{\theta})$.
By definition, we have $w_y=0$.

Formally, Eqs. (\ref{ENSKOG1}), (\ref{FRED1}) look identical to the kinetic equation derived previously in Refs.
\cite{ihle.t:2011,ihle.t.1:2014}
for the regular VM. The difference hides
in the definition of the average angle $\Phi_1$ of the focal particle:
In the regular VM all particles within interaction range are accepted,
irrespective of their orientation. This leads to a straightforward evaluation of the integrals over
the pre-collisional angles $\tilde{\theta}_i$. For the restricted angle model,
the integration domain has to be split into subdomains because the number of angular arguments
in $\Phi_1$ depends on the values of the pre-collisional angles.

For simplicity we assume low densities, $M\ll 1$,
and only include self-interactions and binary collisions.
This amounts to truncating the sum in Eq. (\ref{ENSKOG1}) after $n=2$.
As pointed out in the supplemental material of Ref. \cite{ihle.t:2013}, in order to enforce exact mass conservation,
any such truncation must be accompanied by a consistent rescaling of the Poissonian weight factor.
In our case,
${\rm e}^{-M}$ must be replaced by $1/(1+M)$.
The self-interaction or self-diffusion term with $n=1$ is the same as in the regular VM. However,
in the evaluation of the binary collision term, two cases have to be distinguished,
(i) the direction of particle $2$ deviates too strongly from that of particle $1$,
that is ${\bf v}_2\cdot{\bf v}_1/v_0^2<\rm cos{\,\alpha}$, and (ii) the ``opinion'' of particle $2$ is accepted by particle $1$. In the first case, $\Phi_1=\tilde{\theta}_1$ whereas for case (ii)
$\Phi_1={\rm Arg}[{\rm exp}(i\tilde{\theta}_1)
+{\rm exp}(i\tilde{\theta}_2)]$
which can be reformulated by means of trigonometric identities as
\begin{equation}
\label{TRIGID1}
 \Phi_1(\tilde{\theta}_1,\tilde{\theta}_2) = \begin{dcases*}
        {\tilde{\theta}_1+\tilde{\theta}_2\over 2}     & if $|\tilde{\theta}_1-\tilde{\theta}_2| \leq \pi$\\
        {\tilde{\theta}_1+\tilde{\theta}_2\over 2}+\pi & if $|\tilde{\theta}_1-\tilde{\theta}_2|>\pi$\,.
        \end{dcases*}
\end{equation}

Multiplying Eq. (\ref{ENSKOG1}) by ${\rm cos}(k\theta)$ and integrating over $\theta$ will lead to an infinite set of algebraic equations
for the Fourier coefficients $C_k$,
\begin{equation}
\label{FOURIER_REL1}
C_k={\lambda_k \over 1+M}\left\{\sum_{q=0}^{\infty} A_{kq}\,g_q+2\pi S \sum_{p=0}^{\infty} \sum_{q=0}^{\infty} B_{kpq}(\alpha)\, g_p g_q
\right\}
\end{equation}
where
\begin{equation}
\label{LAMBDA_DEF}
 \lambda_k = \begin{dcases*}
        1     & for $k=0$\\
        {4 \over k\eta}  {\rm sin}\left({\eta k\over 2}\right) & for $k>0$
        \end{dcases*}
\end{equation}
The coupling constants in Eq. (\ref{FOURIER_REL1}) are given by angular integrals,
\begin{eqnarray}
\nonumber
A_{kq}&=&\int_0^{2\pi}{\rm cos}(k\theta_1) {\rm cos}(q\theta_1)\,{d\theta_1\over 2\pi}={1\over 2}\delta_{kq}(1+\delta_{k0}) \\
\nonumber
B_{kpq}(\alpha)&=&
\int_0^{2\pi}
\int_0^{2\pi}{d\theta_1\,d\theta_2\over (2\pi)^2}
{\rm cos}(k\Phi_1)
{\rm cos}(p\theta_1)
{\rm cos}(q\theta_2) \\
\label{ANGLE_INTEGRAL1}
&=&B^{(1)}_{kpq}+B^{(2)}_{kpq}\,,
\end{eqnarray}
where the binary collision couplings depend on the angle $\alpha$ and have been split into two types,
\begin{eqnarray}
\nonumber
B^{(1)}_{kpq}&=&\int_0^{2\pi}{d\theta_1\over 2\pi}{\rm cos}(p\theta_1){\rm cos}(k\theta_1)
\int_{\theta_1+\alpha}^{\theta_1-\alpha+2\pi}{\rm cos}(q\theta_2)\,{d\theta_2\over 2\pi} \\
\nonumber
B^{(2)}_{kpq}&=&\int_0^{2\pi}{d\theta_1\over 2\pi}{\rm cos}(p\theta_1)
\int_{\theta_1-\alpha}^{\theta_1+\alpha}{\rm cos}[k\Phi_1(\theta_1,\theta_2)] \\
&\times& {\rm cos}(q\theta_2)\,{d\theta_2\over 2\pi}
\label{BTYPE}
\end{eqnarray}
Here, $B^{(1)}_{kpq}$ corresponds to the situation where the focal particle rejects its neighbor's ``opinion'',
whereas in $B^{(2)}_{kpq}$ the directions of both particles $1$ and $2$ contribute
to the average direction $\Phi_1$.
The first set of integrals in Eq. (\ref{BTYPE}) is easily evaluated.
For $q>0$ one finds,
\begin{equation}
\label{COUPLING_B1}
B^{(1)}_{kpq}(\alpha)=-{{\rm sin}(\alpha q)\over 4\pi q}
[\delta_{k,p+q}
+\delta_{k,-p-q}
+\delta_{k,p-q}
+\delta_{k,q-p}
]\,,
\end{equation}
where $\delta_{k,q}$ is Kronecker's delta.
For $q=0$ we have,
\begin{equation}
\label{COUPLING_B10}
B^{(1)}_{kp0}={1\over 2}\left(1-{\alpha\over \pi}\right)
[\delta_{k,p}+\delta_{k,-p}]\,.
\end{equation}
In order to transform $B^{(2)}_{kpq}$ into simple trigonometric integrals,
only the first identity of Eq. (\ref{TRIGID1}) is needed because the integrals for $B^{(2)}_{kpq}$ are set up such
that $|\theta_1-\theta_2|<\alpha$ and by definition $\alpha\le\pi$.
This gives
\begin{eqnarray}
\nonumber
B^{(2)}_{kpq}(\alpha)&=&\int_0^{2\pi}{d\theta_1\over 2\pi}{\rm cos}(p\theta_1)
\int_{\theta_1-\alpha}^{\theta_1+\alpha}{\rm cos}[k(\theta_1+\theta_2)/2] \\
\label{COUPLING_B2}
&\times& {\rm cos}(q\theta_2)\,{d\theta_2\over 2\pi}
\end{eqnarray}
The discussion of this integral is relegated to Appendix A.
One major difference of the hierarchy equations, Eq. (\ref{FOURIER_REL1}) to those of the regular VM
is, that the coupling matrices $B_{kpq}$ are asymmetric with respect to interchanging $p$ and $q$.
As discussed in Appendix A, this is due to
the social bias of an agent to favor its own ``opinion''.

To verify the consistency of Eq. (\ref{FOURIER_REL1}), we evaluate the first hierarchy equation for $C_0$.
Mass conservation requires that the homogeneous density $\rho_0$ stays invariant in every iteration,
$\int f(\theta,t+\tau)\,d\theta=\int f(\theta,t)\,d\theta=\rho_0$. Therefore, $C_0$ must be equal to $g_0$.
From Eq. (\ref{ANGLE_INTEGRAL1}) we see that $A_{00}=B_{000}=1$ and that these are the only nonzero coefficients entering
the equation for $C_0$. Substituting these coefficients into Eq. (\ref{FOURIER_REL1}) leads to
\begin{equation}
\label{C0_DEF}
C_0={1\over 1+M}(g_0+2\pi S g_0^2)
\end{equation}
Using Eq. (\ref{G0_DEF}) and the definition of $M=\rho_0 S=\rho_0 \pi R^2$ we obtain the correct result, $C_0=g_0=\rho_0/(2\pi)$.

\begin{figure}
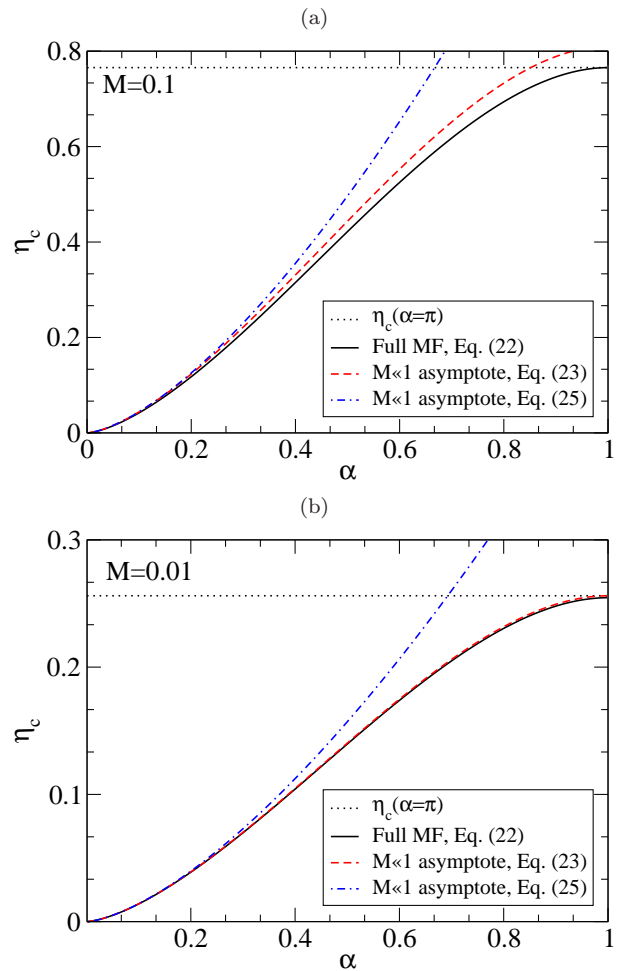

\centering
\subfigure[]{
\includegraphics[width=8cm,clip]{Eq23-26M01.eps}
}
\subfigure[]{
\includegraphics[width=8cm,clip]{Eq23-26M001.eps}
}
\caption{Prediction of the mean-field theory for the critical noise amplitude at different densities (a) $M = 0.1$. (b) $M = 0.01$.
The restriction angle $\alpha$ is given in units of $\pi$.}
\label{fig:mean-field}
\end{figure}

At any noise $\eta$ the hierarchy equations, Eq. (\ref{FOURIER_REL1}), have a trivial solution where all coefficients $g_k=0$ for $k>0$.
This solution describes the disordered state; there is no preference for any direction.
Similar to the regular VM, for low noise we expect an ordered solution with $g_k\neq 0$, that bifurcates off the disordered solution
at the threshold noise $\eta_c$. On this branch of the solution, at a noise slightly below $\eta_c$, the first mode $g_1$ dominates
all higher modes, that is the ratios $g_2/g_1,g_3/g_1,\ldots$ go to zero if $\eta\rightarrow \eta_c$.
Hence, to find the branching point $\eta_c$ we assume stationarity, $C_k=g_k$, and neglect all modes $g_k$ for $k\geq 2$ in Eq. (\ref{FOURIER_REL1}).
Then, the second hierarchy equation yields,
\begin{equation}
g_1={\lambda_1\over 1+M}\left[ A_{11}\,g_1+2\pi S (B_{110}+B_{101})\,g_0 g_1\right]\,,
\end{equation}
and with the help of Eqs. (\ref{G0_DEF}), (\ref{ANGLE_INTEGRAL1}), (\ref{COUPLING_B1}) a
transcendental equation for $\eta_c$ follows:
\begin{eqnarray}
\nonumber
\Lambda_1&=&{2 (1+M\gamma) \over \eta_c(1+M)}\,{\rm sin}\left({\eta_c\over 2}\right)=1 \\
\label{GAMMA_EQ}
\gamma(\alpha)&=&1-{\alpha\over \pi}+{4\over \pi}\sin{ \left( \alpha\over 2\right) } -{1\over \pi}\sin {\left(\alpha\right) }
\end{eqnarray}
where $\Lambda_1$ is the amplification factor for the mode $\tilde{g}_1$, which is proportional to
the $x$-component of the momentum density.
In the limit of no restriction, $\alpha=\pi$, we find $\gamma=4/\pi$
and recover the threshold equation for the regular VM (Eq. (12) in \cite{ihle.t.1:2014} where terms with $n>2$ are truncated).
Expanding Eq. (\ref{GAMMA_EQ}) in the low density limit,
$M\ll 1$, leads to
\begin{equation}
\label{EQ_OLD_24}
\eta_c= \sqrt{24 M\left( \gamma-1\right)}
+O(M)\,.
\end{equation}
For $\alpha=\pi$, this expression agrees with
the small density expansion for the regular VM,  Eq. (13) in \cite{ihle.t.1:2014},
\begin{equation}
\label{ETAC_REGULAR_VM}
\eta_c(\alpha=\pi)= \sqrt{48 M\left( {2\over \pi}-{1\over 2}\right)}
+O(M)\,.
\end{equation}
Furthermore, the function $\gamma-1$ is non-negative and increases monotonically with $\alpha$
for $0\le \alpha \le \pi$, as anticipated.
Investigating the additional limit of strong restriction, Eq. (\ref{GAMMA_EQ}) gives
\begin{eqnarray}
\label{SMALL_ALPHA}
\nonumber
\gamma&=&1+{\alpha^3\over 12\pi}+O(\alpha^5) \\
\eta_c&=&\sqrt{2M/\pi}\,\alpha^{3/2}\;\;\;{\rm for}\;\alpha\ll 1;\;M \ll 1
\end{eqnarray}
Eq. (\ref{SMALL_ALPHA}) predicts that
$\eta_c$ goes to zero at infinite ``ignorance'' $\alpha=0$
where all particles perform independent random walks and never align with anybody.
Thus, as one would intuitively guess, in this case the theory claims that no ordered state exists.
Fig. \ref{fig:mean-field} shows the numerical solution of Eq. (\ref{GAMMA_EQ}) for two different normalized densities, $M=0.1$ and $M=0.01$
and compares it with the asymptotic formulas, Eqs. (\ref{EQ_OLD_24}) and (\ref{SMALL_ALPHA}).
One sees that for small $M\approx 0.01$, the low-$M$ expansion agrees quite well with the exact result for all angles $\alpha$.
The asymptotic power law for $\alpha\rightarrow 0$ is superlinear, $\eta_c\sim \alpha^{3/2}$.
However, due to the change of curvature of the $\eta_c(\alpha)$-curve from positive to negative, at intermediate angles
$\alpha\approx 0.2\pi\ldots 0.6\pi$ it appears as if there is linear scaling, $\eta_c\sim \alpha$.
As discussed further below, this is what we observed in agent-based simulation which were not performed for very small angles,
$\alpha<0.2\pi$, due to numerical limitations.

\subsection{Analytical solution and mean-field tricritical point for polar order}
\label{sec:anal}

Near the flocking threshold, higher order Fourier modes are suppressed, $g_1\gg g_2 \gg g_3\ldots$
and Eq. (\ref{FRED1}) can be straightforwardly solved by
setting higher modes to zero, truncating the infinite hierarchy (\ref{FOURIER_REL1})
after the first few equations.
The Fourier coefficients are normalized by means of the density $\rho_0$,
\begin{equation}
\label{NORMALIZE}
\tilde{g}_k=2\pi {g_k\over \rho_0}
\end{equation}
which amounts to the choice $\tilde{g}_0=1=const$.
Then, the first three nontrivial hierarchy equations from Eq. (\ref{FOURIER_REL1}) become,
\begin{eqnarray}
\nonumber
\tilde{g}_1&=&\alpha_1(\tilde{g}_1+4M[\bar{B}_{101}\tilde{g}_1+\bar{B}_{112}\tilde{g}_1\tilde{g}_2
+\bar{B}_{123}\tilde{g}_2\tilde{g}_3]) \\
\nonumber
\tilde{g}_2&=&\alpha_2(\tilde{g}_2+2M[\bar{B}_{211}\tilde{g}_1^2+2\bar{B}_{202}\tilde{g}_2+2\bar{B}_{213}\tilde{g}_1\tilde{g}_3]) \\
\label{FIRST_PART}
\tilde{g}_3&=&\alpha_3(\tilde{g}_3+4M[\bar{B}_{303}\tilde{g}_3+\bar{B}_{312}\tilde{g}_1\tilde{g}_2])\\
\label{G3_HIERARCH}
\alpha_k&=&{2\over k \eta(1+M)}{\rm sin}\left({\eta k\over2}\right)\,.
\end{eqnarray}
Here, we kept only the modes $\tilde{g}_0$ to $\tilde{g}_3$ and neglected all others
because we are mainly interested in understanding the nature of the bifurcation to a (homogeneous) ordered state.
The coupling constants $\bar{B}_{knm}$ are obtained by symmetrization of the coefficients defined in Eq. (\ref{ANGLE_INTEGRAL1}),
\begin{equation}
\label{SYMMETRIZE}
\bar{B}_{knm}={1\over 2}(B_{knm}+B_{kmn})
\end{equation}
The $\bar{B}_{knm}$ depend on the restriction angle $\alpha$ and are given in Appendix A.
Starting from the third line, equations (\ref{FIRST_PART})
can be solved successively leading to
expressions of
the higher modes in terms of the first mode $\tilde{g}_1$,
\begin{eqnarray}
\nonumber
\tilde{g}_2={n_2\over 1-[a_2+b_2\tilde{g}_1^2]}\tilde{g}_1^2 \\
\label{G-EQUS}
\tilde{g}_3={n_2n_3\over 1-[a_2+b_2\tilde{g}_1^2]}\tilde{g}_1^3
\end{eqnarray}
with the abbreviations
\begin{eqnarray}
\nonumber
n_2&=&2M\alpha_2\bar{B}_{211} \\
\nonumber
n_3&=&{4M\alpha_3\bar{B}_{312}\over 1-\alpha_3[1+4M\bar{B}_{303}]} \\
\nonumber
a_2&=&\alpha_2(1+4M\bar{B}_{202})\\
b_2&=&4\alpha_2M\bar{B}_{231}n_3
\end{eqnarray}
Substituting Eqs. (\ref{G-EQUS}) into the first line of Eq. (\ref{FIRST_PART})
yields a closed expression for the first mode $\tilde{g}_1$,
\begin{equation}
\label{CLOSED_TRIC}
1=\Lambda_1+D_2(\tilde{g}_1^2)\tilde{g}_1^2+D_4(\tilde{g}_1^2)\tilde{g}_1^4
\end{equation}
with $\Lambda_1$ defined in Eq. (\ref{GAMMA_EQ}) and
\begin{eqnarray}
D_2&=&{4M \alpha_1 n_2\bar{B}_{112}\over 1-[a_2+b_2\tilde{g}_1^2]} \\
D_4&=&{4M \alpha_1 n_2^2n_3\bar{B}_{123}\over (1-[a_2+b_2\tilde{g}_1^2])^2}
\end{eqnarray}
where at the threshold $\eta=\eta_c(M)$ one has $\Lambda_1=1$ (see Eq. (\ref{GAMMA_EQ})), and $\tilde{g}_1=0$.
The character of the bifurcation to the ordered state, that is to non-zero $\tilde{g}_1$, depends on the sign
of the coefficient $D_2$, evaluated at the threshold.
Therefore, the condition for a tricritical point, where the character changes from subcritical to supercritical is
\begin{equation}
\label{D2_COND}
D_2(\eta=\eta_c,\tilde{g}_1=0)=0
\end{equation}
This is only possible if at least one of the following quantities vanish, $\alpha_1$, $\alpha_2$, $\bar{B}_{211}$ or $\bar{B}_{112}$. In the low density limit that our approach is based on, the threshold noise is small, $\eta_c\sim\sqrt{M}\ll 1$, and thus
$\alpha_1$ and $\alpha_2$ are of order one and cannot be zero. Furthermore, it is easy to see from Eq. (\ref{APPENDIX_BLAST})
that $\bar{B}_{211}$ can only be zero in the limit $\alpha=0$ where no ordered state exists.
However, the equation $\bar{B}_{112}=0$ has a solution at the angle
$\alpha_c=0.4429096\,\pi$.
Thus, assuming spatially homogeneous states, we found an apparent tricritical point:
for all angles $\alpha$ smaller than $\alpha_c$ the flocking transition
is discontinuous, in the mean-field limit of large mean free path. For angles larger than $\alpha_c$ and in a small system, the transition
is predicted to be continuous. However, one has to keep in mind that
in the regular VM, the homogeneous flocking state has a long-wave instability right next to the flocking threshold \cite{bertin.e:2009,ihle.t:2011}.
This leads to inhomogeneous, soliton-like states that change the order of the
phase transition to discontinuous in systems larger than a critical size $L_c$ \cite{ihle.t:2013}, even if the transition of a homogeneous system is predicted to be continuous.
Something similar is expected in our model and will be investigated further below.
For a justification of the term ``phase transition'' in finite systems, see Appendix B.

Note that the inhomogeneous
reorganization of the system due to emerging soliton-like waves at large system sizes is
qualitatively different from the well-known finite-size effects of equilibrium systems in the (grand)canonical ensemble.
In these systems, the trivial
fact that the correlation length cannot exceed the system size is used to set up
finite-size scaling and to extract the behavior at infinite system size.
As shown in Ref. \cite{baglietto.g:2008} it is possible to perform such a finite-size scaling
analysis of the VM at system sizes below $L_c$ and to
obtain critical exponents as well as consistent hyper-scaling relations for a second-order
phase transition. However, 
unless $L_c$ is infinite for the particular model and the parameters used,
such an analysis would not
extract the correct behavior in the thermodynamic limit because it would miss possible density instabilities.

\begin{figure}
\centering
\includegraphics[width=8cm,clip]{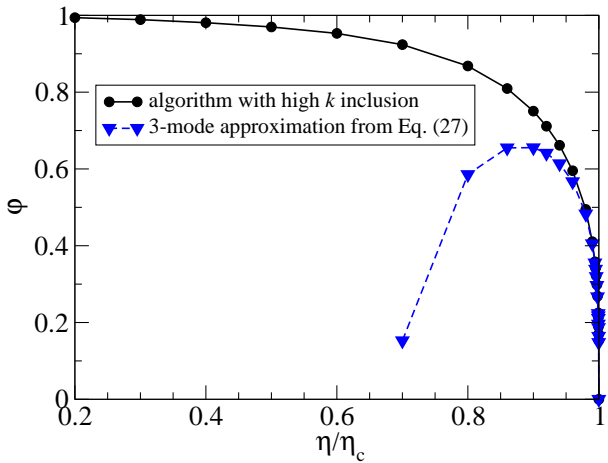}
\caption{Comparison of the scaling of the order parameter $\varphi=\tilde{g}_1/2$ with noise
$\eta$ obtained analytically using Eq. (\ref{FIRST_PART}) (all modes higher than $k=3$ are set to zero) and with the high-$k$ approximation method described in section \ref{sec:fredholm}.
Lines only serve as guides to the eye.
Here, $\alpha=\pi$, and $k_{max}=30$.}
\label{fig:order_theory}
\end{figure}

It is interesting to see that (i) the tricritical angle $\alpha_c$
does not depend on density, at least in the low density limit considered here, and (ii) its mathematical cause is the vanishing
of the coupling between the modes $\tilde{g}_1$ and $\tilde{g}_2$ in the hierarchy equation
for $\tilde{g}_1$.
Analyzing Eq. (\ref{CLOSED_TRIC}) at the tricritical point $\eta=\eta_c$ and $\alpha=\alpha_c$ leads to the
mean-field
exponent of $1/4$ for the order parameter scaling,
\begin{eqnarray}
\nonumber
\tilde{g}_1\sim (\eta_c-\eta)^{1/4}\;\;{\rm for}\;\alpha=\alpha_c,\,\eta \lessapprox \eta_c\\
\label{SCALING_MF}
\tilde{g}_1\sim (\eta_c-\eta)^{1/2}\;\;{\rm for}\;\alpha>\alpha_c,\,\eta \lessapprox \eta_c
\end{eqnarray}
The structure of the denominator of the coefficients $D_2$ and $D_4$ also provides an estimate on the validity of the
three-mode expansion. If the distance $\eta_c-\eta$ to the threshold noise is increased, the order parameter $\tilde{g}_1$
grows. Once it is so large that $1-[a_2+b_2\tilde{g}_1^2]$ goes to zero, the approximation is expected to break down violently. This
already happens not too far from the threshold and can be seen in Fig. \ref{fig:order_theory}, where the blue
dashed curve describes the analytical solution of Eq. (\ref{FIRST_PART}).
This approximation neglects all modes $\tilde{g}_k$ with $k$ larger than three.
Near the threshold, as anticipated, it agrees perfectly with the numerical solution of the Fredholm equation which is explained in a
later chapter.
However, deeper in the ordered phase, at smaller noise, the 3-mode-approximation
starts to show unphysical behavior.

The physical reason for this deviation is the negligence of the higher modes $\tilde{g}_4,\tilde{g}_5,\ldots$,
which in reality would start to grow
when $\tilde{g}_1$ increases.
In fact, using the hierarchy equations (\ref{FOURIER_REL1}) it can be shown that for $\alpha=\pi$ and $\eta\rightarrow 0$
all modes $\tilde{g}_k$ for $k>0$ become equal to the same value $\tilde{g}_k=2$.
This is a simple consequence of the fact that at zero noise all particles take on the same orientation
and the distribution function $f(\theta)$ becomes equal to the periodically-continued delta-function,
$\hat{\delta}(\theta-\theta_0)$.

\subsection{Nematic solutions and tricritical points}
\label{sec:nematic}
\begin{figure}
\centering
\includegraphics[width=8.0cm,clip]{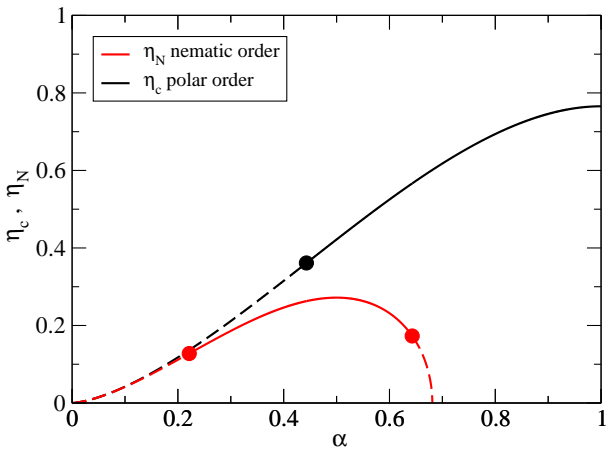}
\caption{The critical noises for the transition from disorder to polar order (black curves), $\eta_c$, and
to nematic order (red curves),
$\eta_N$,
are plotted as a function of restriction angle $\alpha$.
The curves were calculated by means of Eqs. (\ref{GAMMA_EQ}) and (\ref{THRESH_NEM}), respectively.
The circles denote tricritical points at which the transition changes from discontinuous (dashed lines) to continuous (solid lines).
}
\label{FIG_NEMATIC}
\end{figure}

As described in section \ref{results}, in agent-based simulations we sometimes encounter
groups of particles which move in opposite directions. These states are characterized by a
large nematic order parameter $Q$ but only a small polar order parameter $\varphi$.
In social science's models of opinion dynamics, this is called \emph{polarization}
\cite{hegselmann.r:2002,kurmyshev.e:2011,baronchelli.a:2007}.
To analytically explore such a possibility
we reanalyze the infinite hierarchy for the Fourier coefficients, Eq. (\ref{FOURIER_REL1}) with respect to nematic order.
Compared to a state of polar order, a perfect nematic state
has the additional symmetry of $f(\theta\pm \pi)=f(\theta)$.
This requires all odd Fourier coefficients in the series, Eq. (\ref{FOURIER1}), and therefore also the polar order
parameter $\varphi$, to vanish exactly.
Setting $g_1=g_3=\ldots = g_{2k+1}=0$ in Eq. (\ref{FOURIER_REL1}), one realizes that the
odd and even coefficients are decoupled: while all odd coefficients $C_1$, $C_3,\ldots$ become zero,
the even coefficients $C_{2k}$ can be non-zero and depend only on $g_{2n}$.
This is because the coefficients $A_{kq}$ and $B_{kpq}$ always
vanish if neither $|k| = |p+q|$ nor $|k|=|p-q|$ is true.
Setting $C_{2k}=g_{2k}$ and normalizing the coefficients as prescribed by Eq. (\ref{NORMALIZE}),
we obtain a new hierarchy for stationary nematic states,
\begin{eqnarray}
\nonumber
\tilde{g}_2&=&\alpha_2(\tilde{g}_2+4M[\bar{B}_{202}\tilde{g}_2+
\bar{B}_{224}\tilde{g}_2\tilde{g}_4+\bar{B}_{246}\tilde{g}_4\tilde{g}_6]) \\
\nonumber
\tilde{g}_4&=&\alpha_4(\tilde{g}_4+2M[\bar{B}_{422}\tilde{g}_2^2+
2\bar{B}_{404}\tilde{g}_4+2\bar{B}_{426}\tilde{g}_2\tilde{g}_6]) \\
\label{NEMATIC_HIER}
\tilde{g}_6&=&\alpha_6(\tilde{g}_6+4M[\bar{B}_{606}\tilde{g}_6+\bar{B}_{624}\tilde{g}_2\tilde{g}_4])
\end{eqnarray}
where modes $\tilde{g}_8$ and higher are neglected.
The coupling coefficients $\bar{B}_{kpq}$ are given in Appendix A.
Similar to the procedure for finding polar order, we check whether
a nematic state can bifurcate from a disordered state at some critical noise value $\eta=\eta_N$.
Near such a (so far hypothetical) bifurcation, all modes higher than $\tilde{g_2}$ are negligible
and the first equation of the hierarchy, Eq. (\ref{NEMATIC_HIER}), gives the following consistency condition,
\begin{equation}
\label{THRESH_NEM}
\Lambda_2=\alpha_2(1+4M \bar{B}_{202})=1
\end{equation}
where $\Lambda_2$ is the amplification factor of the mode $\tilde{g}_2$.
Interestingly,
while this transcendental equation has no nontrivial solutions for the regular Vicsek model at any density,
it does have a solution $\eta_N>0$ if the restriction value $\alpha$ is smaller than a cut-off angle
$\alpha_N=0.68092\,\pi$.
Fig. \ref{FIG_NEMATIC} shows a plot of the nematic threshold noise $\eta_N$ as a function of $\alpha$ together with
the polar threshold noise $\eta_c$ at density $M=0.1$. The cut-off $\alpha_N$ does not depend on particle density,
at least in the low density approximation applied here.
Fig. \ref{FIG_NEMATIC} tells us that at small enough $\alpha$ and $\eta$, stationary nematic states do exist.
We also see that at small $\alpha$,
the threshold noise for nematic states is only slightly below the threshold for polar states, whereas the relative
difference $(\eta_c-\eta_N)/\eta_c$ goes to one if the cut-off $\alpha_N$ is approached from below.
While our analytical
discovery of stationary nematic states does not guarantee that these states are stable and relevant for the long-time behavior of agent-based simulations, our results are consistent with
the apparent longevity of states with large nematic but small polar order in microscopic simulations.
Of course, to fully explore the competition of polar, nematic and other apolarly ordered states,
a comprehensive stability analysis similar to Refs. \cite{chou.yl:2012,menzel.a:2012},
and more simulations are needed. This is beyond the scope of this paper.
However, similar to the case of polar order, it is straightforward to
analyze whether the bifurcation from a disordered to a (homogeneous) nematic state is continuous or discontinuous.
For this purpose, we successively solve the hierarchy equations, Eq. (\ref{NEMATIC_HIER}),
by first expressing the higher modes $\tilde{g}_4$ and $\tilde{g}_6$ in terms of $\tilde{g}_2$:
\begin{eqnarray}
\tilde{g}_4&=&{n_4\over 1-[a_4+b_4 \tilde{g}_2^2]} \tilde{g}_2^2 \\
\tilde{g}_6&=&n_6 \tilde{g}_2 \tilde{g}_4={n_4 n_6\over 1-[a_4+b_4 \tilde{g}_2^2]} \tilde{g}_2^4
\end{eqnarray}
with the abbreviations
\begin{eqnarray}
\nonumber
a_4&=&\alpha_4(1+4 M \bar{B}_{404})\\
\nonumber
a_6&=&\alpha_6(1+4 M \bar{B}_{606})\\
\nonumber
n_4&=&2\alpha_4 M \bar{B}_{422}\\
\nonumber
n_6&=&{4 \alpha_6 M \bar{B}_{624}\over 1-a_6} \\
b_4&=&4M n_6 \bar{B}_{426}\:
\end{eqnarray}
Inserting these expressions into the first hierarchy equation, Eq. (\ref{NEMATIC_HIER}), leads to a closed-form expression for $\tilde{g}_2$,
\begin{equation}
\label{CLOSED_NEMA}
1=\Lambda_2+H_2(\tilde{g}_2^2)\,\tilde{g}_2^2+
  H_4(\tilde{g}_2^2)\,\tilde{g}_2^4
\end{equation}
with
\begin{eqnarray}
H_2&=&{4M \alpha_2 n_4 \bar{B}_{224}\over
1-[a_4+b_4\tilde{g}_2^2]}\\
H_4&=&{4M\alpha_2 n_4^2 n_6 \bar{B}_{246}\over (1-[a_4+b_4 \tilde{g}_2^2])^2}
\end{eqnarray}
One has $\Lambda_2(\eta_N)=1$, and
similar to the polar case, see Eq. (\ref{D2_COND}),
the zeros of the quadratic coefficient $H_2$ define tricritical points,
\begin{equation}
\label{TRIC_NEMA}
H_2(\eta=\eta_N,\tilde{g}_2=0)=0\,,
\end{equation}
where the bifurcation changes from subcritical to supercritical or vice versa.
Surprisingly, below the cut-off angle $\alpha_N$ we find two tricritical points,
$\alpha_{c,1}=0.22145\,\pi$ and $\alpha_{c,2}=0.64299\,\pi$.
Further analyzing $H_2$ and $H_4$ at the critical line $\eta=\eta_N(\alpha)$, we find that $H_2>0$ and $H_4<0$ for
both $\alpha<\alpha_{c,1}$ and $\alpha>\alpha_{c,2}$.
Thus, the transition would be discontinuous in these two regions. Note, that $1-\Lambda_2$ is always negative for $\eta<\eta_N$.
In Fig. \ref{FIG_NEMATIC}, discontinuous bifurcations are described by dashed lines. Inbetween the two points, e.g. at $\alpha_{c,1}
<\alpha<\alpha_{c,2}$,
$H_2$ is positive and, depending on $\alpha$, $H_4$ can be either positive or negative.
This means that the transition to a homogeneous nematic state from the disordered state is continuous in this middle section,
denoted by a solid red line in Fig. \ref{FIG_NEMATIC}.

\subsection{Fragmentation into ordered groups}
\label{sec:fragment}
\begin{figure}
\centering
\includegraphics[width=8.0cm,clip]{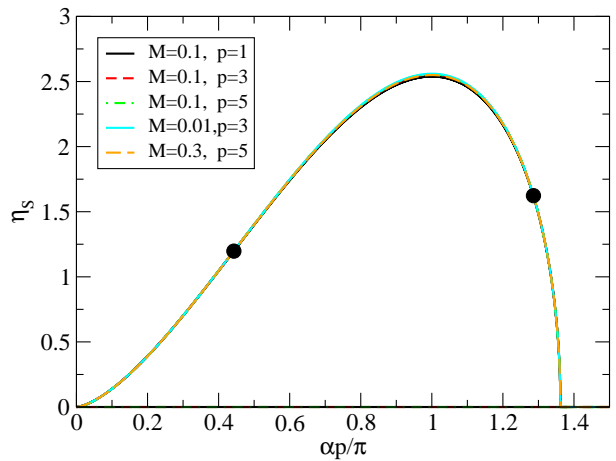}
\caption{The scaled critical noises $\eta_S=
\eta_c^{(p)}p^{3/2}\sqrt{1+M\over M}$ for mean-field transitions from the disordered state to symmetric states with $p$ fragments
versus the scaled angle $\alpha p/\pi$ for a variety of densities $M$ and fragment numbers $p$.
The filled circles represent tricritical points where the character of the transition changes from discontinuous to continuous or vice versa.
}
\label{FRAG_PLOT}
\end{figure}
Models of bounded confidence in social science can have
states that consist of several clusters of opinions
in which agents achieve local consensus
\cite{hegselmann.r:2002,kurmyshev.e:2011,baronchelli.a:2007}.
This is called {\em opinion fragmentation}.
Translated into the language of our system, fragmentation would correspond to several groups of particles whose
members interact with each other in every iteration because their angular differences are smaller than $\alpha$,
but there is hardly any interaction between members of different groups.
The state of polar order would then correspond to {\em global consensus} where only one group exists.
The nematic state would be an example for
fragmentation into two groups. A natural question to ask is whether our model also allows states with three, four and more distinct groups.
It turns out
that the analysis of the hierarchy equation, Eq. (\ref{FOURIER_REL1}),
for polar and nematic order can be straightforwardly generalized to fragmentation into $p=3,4,5,\ldots$ groups.
In this paper, we will restrict ourselves to highly symmetrical arrangements where the mean angles of the participating
groups point into the directions $\bar{\theta}_k=2\pi k/p$ with $k=0,1,\ldots p-1$.
In these arrangements, the distribution function $f$ has the ``mirror''-symmetry $f(\bar{\theta}_k+\theta)=
f(\bar{\theta}_k-\theta)$. This restriction allows us to still use the Fourier cosine expansion, Eq. (\ref{FOURIER1}).
The more general case requires the inclusion of sine terms and will be left for the future.
Symmetric fragmentation into $p$ groups requires the vanishing of all coefficients $g_k$ whose mode numbers are not multiples of $p$ due
to the $p$-fold symmetry, $f(\theta+2\pi k/p)=f(\theta)$. 
For example, at $p=3$, three groups move into the three main directions $0,2\pi/3$ and $4\pi/3$, and are described by the
Fourier coefficients $g_0,g_3,g_6,g_9,\ldots$. Similar to the nematic case with $p=2$,
these coefficients only couple to themselves, e.g. they do not generate
coefficients such as $g_1$ or $g_4$ which are supposed to remain zero.
All expressions and coefficients for the nematic state
can be easily generalized by formal replacements such as $g_2\rightarrow g_p$,
$g_4\rightarrow g_{2p}$,
$g_6\rightarrow g_{3p}$, $n_4\rightarrow n_{2p}$, $\bar{B}_{426}\rightarrow \bar{B}_{2p,p,3p}$ and so on.
In particular, the threshold noise $\eta_c^{(p)}$
for the bifurcation of the disordered state to a state with $p$ symmetric fragments follows from the condition that the amplification rate $\Lambda_p$ for
the mode $\tilde{g}_p$ is equal to one,
\begin{equation}
\label{LAMBDA_P_DEF}
\Lambda_p=\alpha_p(1+4 M \bar{B}_{p0p})
\end{equation}
where
\begin{equation}
\label{BP0P_DEF}
\bar{B}_{p0p}={1\over 4}\left(1-{\alpha  \over \pi} \right)+
{{\rm sin}\left({p \alpha\over 2 }\right)\over \pi p}
-{{\rm sin}\left(p\alpha\right)\over 4 \pi p}\,.
\end{equation}
Setting $\eta=0$ in Eq. (\ref{LAMBDA_P_DEF}),
the cut-off selectivity angle $\alpha_0^{(p)}$ above which stationary fragmented states are impossible,
follows from the transcendental equation
\begin{equation}
\label{CUTOFF_P}
\pi z+{\rm sin}(\pi z)-4 {\rm sin}\left({\pi z\over 2}\right)=0
\end{equation}
with $z=\alpha_0^{(p)} p/\pi$.
The solution is $z=1.361845$, and therefore the cut-off for a fragmented state of $p$ main directions is
$\alpha_0^{(p)}=1.361845\,\pi/ p$. The scaling $\sim 1/p$ has a simple physical interpretation:
The difference between the mean angles of adjacent groups $\delta=\bar{\theta}_{k+1}-\bar{\theta}_{k}$ is equal to $2\pi/p$.
As long as $\alpha$ is smaller than this difference $\delta$, members of the two groups have only a negligible chance of interaction. It is plausible to assume that the maximum possible restriction angle scales with the available angular range $\delta$, $\alpha_0^{(p)}\sim \delta\sim 1/p$, which is indeed what we find analytically.
The ratio $\alpha_0^{(p)}/\delta=0.68092$ means that if $\alpha$ is smaller
than $68.1\%$ of the ``opinion'' difference $\delta$, fragmentation becomes possible.
In this notation, the critical noises for the transition from disorder to polar or nematic order
are just special cases with $\eta_c\equiv\eta_c^{(1)}$ and $\eta_N\equiv\eta_c^{(2)}$.

Analyzing Eqs. (\ref{LAMBDA_P_DEF}) and (\ref{BP0P_DEF}),
we find that the critical noises for all possible symmetric fragmentations with $p=1,2,\ldots$ follow an approximate scaling law
\begin{equation}
\label{FRAG_SCALE}
\eta_c^{(p)}=\sqrt{{M\over 1+M}}\; p^{-3/2}\, \Psi\left({\alpha p\over \pi}\right)
\end{equation}
with the universal scaling function $\Psi(z)$.
According to relation (\ref{FRAG_SCALE}), plotting the scaled critical noise $\eta_S=\eta_c^{(p)} p^{3/2} \sqrt{{1+M\over M}}$
as a function of $z= p \alpha/\pi$ should lead to a single curve for different values of $M$ and $p$.
This is indeed what is seen in Fig. \ref{FRAG_PLOT}.
Even though the scaling law is supposed to be only asymptotically valid in the limit $p\rightarrow \infty$ and $M\rightarrow 0$,
the figure shows that it is very accurate even at $p=1$ and $M=0.3$.
The scaling function has the following properties: $\Psi(z)$ is zero for $z\ge 1.361845$ as a consequence of Eq. (\ref{CUTOFF_P}).
In agreement with the result for $p=1$ and small $\alpha$, Eq. (\ref{SMALL_ALPHA}),
the small argument behavior of the scaling function is given by
\begin{equation}
\Psi(z)=\pi \sqrt{2}\,\,z^{3/2}\,\;\;\;\;{\rm for}\;\;z\ll 1
\end{equation}
From Fig. \ref{FRAG_PLOT} one can also read off the value $z_{max}=1$ where $\Psi$ has its maximum.
Therefore, the critical noises of the different fragmented states take their largest value of $2.555\sqrt{{M/(1+M)}}p^{-3/2}$
at $\alpha=\pi/p$.

Finally, the calculation of tricritical points for polar and nematic states can be generalized to fragmented states.
Mathematically, the zeros of the coupling coefficient $\bar{B}_{p,p,2p}$,
\begin{equation}
\label{bpp2p}
\bar{B}_{p,p,2p}=-{{\rm sin}( 2 p \alpha)\over 16 \pi p}
                 -{{\rm sin}( p \alpha)\over 8 \pi p}
                 +{1\over 6 \pi p} {\rm sin}\left( {3 p \alpha\over 2}\right)
\end{equation}
determine the tricritical points.
This amounts to solving the equation,
\begin{equation}
\label{TRIC_P}
-3{\rm sin}( 2 \pi z)
-6{\rm sin}( \pi z)
+8{\rm sin}\left( {3 \pi z\over 2}\right)=0\,
\end{equation}
with $z= p\alpha/\pi$.
Apart from the trivial solution $z=0$, Eq. (\ref{TRIC_P}) has two solutions that give the two
tricritical points $\alpha_{c,1}(p)=0.44291\pi/p$ and $\alpha_{c,2}(p)=1.28598\pi/p$.
For the case of polar order, $p=1$, the first tricritical point agrees with the one calculated earlier in chapter \ref{sec:anal}.
The second solution, $\alpha_{c,2}(p=1)$ is physically irrelevant because it is larger than the maximum possible restriction angle of $\pi$.
However, for nematic and all higher (symmetrically) fragmented states,
the second tricritical point is {\em below} the cut-off angle
$1.361845 \pi/p$, and therefore should have physical relevance.
In summary, for $p=1$ we find one tricritical point, whereas for $p\ge 2$ we always have two tricritical points.
In the scaled plot, Fig. \ref{FRAG_PLOT}, the tricritical points for different values of $p$ and $M$ end up on top of each other,
and are given by filled circles.

\subsection{A numerical method for the Fredholm equation}
\label{sec:fredholm}

As we saw in section \ref{sec:anal}, while solutions based on truncating the hierarchy equations for $\tilde{g}_k$
after
the first few modes give good results near the flocking threshold, they will inevitably break down at small enough noise.
Even if one stays away from too small noises, one still would have to solve a large set of equations with a huge number of
mode coupling terms.
We did not succeed in finding a numerical method in the mathematical literature, that is able to solve the singular
nonlinear Fredholm equation (\ref{FRED1}) at high accuracy in the small noise case.
One could argue that a Fourier representation is not suited here and that another set of base functions
might be more appropriate.
Instead of searching for such a set, we decided to stick with the
Fourier representation and to exploit the exactly known solution at $\eta=0$. In other words, we set up
a ``low-temperature'' expansion which is constructed to become exact at zero noise $\eta$ \cite{FOOT3.ihle:2014,lobaskin.v:2013}.
The key idea is to keep the lowest Fourier modes explicitly and to not just neglect higher modes but
to treat them in an approximate fashion. This is related in spirit to the first step of dynamic renormalization \cite{ma.sk:1975},
where equations for higher modes
are approximately solved and expressed in terms of lower modes. These higher modes will then renormalize the lower modes.
Here, we first split the angular mode space into a lower part, $0\le k \le k_1$ and a higher part $k_1 < k \le k_{max}$.
All modes higher than $k_{max}$ are neglected.
The modes of the lower part are treated explicitly and follow the first $k_1$ equations of the hierarchy, Eq. (\ref{FOURIER_REL1}),
with $k_1$ being relatively small, $k_1=3\ldots 5$.
The short wavelength cut-off is chosen very large, $k_{max}\ge 500$, and the Fourier coefficients in the higher mode range
are modeled
by a geometric series,
\begin{equation}
\label{GEOM1}
g_k=g_{k_1} \mu^{(k-k_1)}\;\;\;\;{\rm for}\; k\ge k_1.
\end{equation}
The decay factor $|\mu|<1$ is determined self-consistently from the current ratio of the last two explicitly calculated
modes, $\mu=g_{k_1}/g_{k_1-1}$.
For very small noise, this ratio is always positive and approaches one.
In general, the Fourier modes do not follow a geometric series
unless at $\eta=0$, where $\mu=1$. That means, on the one hand, our approach becomes exact in this perfect-order
limit (and if $k_{max}$ is sent to infinity).
On the other hand, the approach also becomes exact at the threshold, where the higher Fourier modes are irrelevant and the geometric series
assumption does not affect the results.
An approximation that becomes exact at two limits $\eta=0$ and $\eta=\eta_c$ is very likely to only show tiny errors in between.
By increasing both $k_1$ and $k_{max}$ these errors can be further reduced systematically.
To make this idea practically applicable, one more problem has to be solved:
The first $k_1$ equations in Eq. (\ref{FOURIER_REL1}) for the explicit calculation of $g_k$ contain a huge
number of terms because these modes
couple to the $\approx 500$ modes from the higher mode space.
To make things worse, the coefficients of these coupling terms are integrals of the type $\bar{B}_{kpq}$,
which all would have to be calculated beforehand.
As a consequence, solving even the first three to five hierarchy equations would become time-prohibitive.
The way out of this computational disaster is to
rewrite the hierarchy equations by evaluating the binary collision part of the collision operator in real space.
This works as follows: Suppose the first $k_1$ modes are known. Then the modes for $k_1<k\le k_{max}$ are quickly calculated by the geometric series formula, Eq. (\ref{GEOM1}).
Now, we go back to the real space representation of the distribution $f$, given by the inverse Fourier transformation rule,
Eq. (\ref{FOURIER1}).
Using dimensionless distributions and coefficients,
\begin{eqnarray}
\nonumber
\tilde{f}&=&2\pi {f\over \rho_0} \\
\label{FOURIER1_DIM}
\tilde{C}_k&=&2\pi {C_k\over \rho_0}\,,
\end{eqnarray}
Eq. (\ref{FOURIER1}) becomes,
\begin{equation}
\label{NEWSUM}
\tilde{f}(\theta)=1+\sum_{k=1}^{k_1}\tilde{g}_k\,{\rm cos}(k\theta)+\tilde{g}_{k_1}\sum_{q=k_1+1}^{k_{max}}\mu^{(q-k_1)}\,{\rm cos}(q\theta)
\end{equation}
Even though there is about $k_{max}\approx 500$ terms to add, evaluating Eq. (\ref{NEWSUM}) is still quite fast.
One can even go one step further and sum up the higher mode part analytically while sending $k_{max}$ to infinity.
Rewriting
the last sum in Eq. (\ref{NEWSUM}) as two geometric sums
by using the identity ${\rm cos}(k\theta)=({\rm e}^{ik\theta}+{\rm e}^{-ik\theta})/2$ gives,
\begin{eqnarray}
\label{SUMMED_UP}
& &\sum_{q=k_1+1}^{\infty}\mu^{(q-k_1)}\,{\rm cos}(q\theta)= \\
\nonumber
& &{\mu\big[
(1-\mu\cos{\theta})\cos{((k_1+1)\theta)}-\mu \sin{\theta} \sin{((k_1+1)\theta)}
\big]
\over (1-\mu\cos{\theta})^2+\mu^2\sin^2{\theta}}
\end{eqnarray}
Then, $\tilde{f}(\theta)$ is used to evaluate the dimensionless Fourier modes $\tilde{C}_k$ of the
collision integral $I[f]$,
\begin{eqnarray}
\nonumber
& &\tilde{C}_k=\alpha_k\Big[\tilde{g}_k \\
& &+2M\int_0^{2\pi} {d\tilde{\theta}_1\over 2\pi} \int_0^{2\pi} {d\tilde{\theta}_2\over 2\pi}
\tilde{f}(\tilde{\theta}_1)\,\tilde{f}(\tilde{\theta}_2)\,{\rm cos}(k\Phi_1)
\Big]
\label{HALF_REAL}
\end{eqnarray}
This expression is equivalent to the hierarchy equations, Eq. (\ref{FOURIER_REL1}).
The difference is that the part of $I[f]$
that describes binary collisions is now expressed in real space.
The two-dimensional integral in Eq. (\ref{HALF_REAL})
is evaluated numerically for $k=1,\ldots,k_1$ by the
Trapezoidal rule using equidistant angular points.
Thus, we have obtained the first $k_1$ Fourier modes of $\tilde{f}(\theta,t+\tau)$.
In order to solve the fixed point equation for the stationary solution iteratively, we
take the obtained modes as an input for the next iteration, that is set $\tilde{g}_k=\tilde{C}_k$.
We find that this iterative procedure always converges at arbitrary noise $\eta>0$,
probably because it amounts to following
the physically correct time-dependent behavior of $f$ in a very small system with periodic boundary conditions.
Thus, the mathematical procedure is not artificial but reflects physical reality.
Furthermore, switching between real space and Fourier
space representations allows us to ensure
that the zeroth mode $f_0$ always stays constant and, hence, mass is exactly conserved.
This eliminates a source of possible divergence.
The effective ``filtering'' by gently forcing the higher modes to decay geometrically
might be another reason for the robust convergence behavior.

Even though Eq. (\ref{SUMMED_UP}) formally allows us to choose $k_{max}=\infty$,
the current algorithm still does not work exactly at $\eta=0$ but one can get very close to zero noise.
In our implementation, accurate results were obtained down to $\eta=0.02$.
The reason is simply that the discretization of the two-dimensional angular integral in Eq. (\ref{HALF_REAL}) provides
an implicit restriction: at $\eta=0$ the distribution function $f$ becomes equal to the periodically continued Dirac-delta
function, which cannot be accurately resolved on a discrete lattice.
This means the smaller the noise the more discretization points have to be used.

Figures \ref{fig:order_theory}, \ref{fig:vra_polar_lowalpha}(b) and \ref{fig:vra_polar_largealpha}(b)
show results for the order parameter $\varphi=\tilde{g}_1/2$ obtained by the algorithm described above.
A criteria about the quality of the algorithm is whether the order parameter extrapolates to the value $\varphi=1$
in the limit $\eta\rightarrow 0$. This is indeed the case for the 500-mode numerical approach used in Fig. \ref{fig:vra_polar_largealpha}(b).
The curve also shows excellent agreement to agent-based simulations at all noise values.
In Fig. \ref{fig:order_theory} one also sees perfect agreement with the three-mode analytical solution, Eq. (\ref{FIRST_PART}), near the threshold.

\subsection{Simulation}
\label{sec:simulation}

In our two-dimensional model, $N$ point particles move in a rectangular simulation box of size $L_x \times L_y$ with periodic boundary conditions, so that the average particle number density is given by $\rho_0 = N/(L_x L_y)$. The direction of motion of each particle is modified by aligning interactions with other particles located at distance equal or less than $R$ and with the velocities within the angle $\alpha$ from its own velocity. The positions of particles in our simulations are updated by streaming along to new direction according to
%
%
the standard Vicsek updating scheme, given by Eq. (\ref{VM_UPDATE}) with the forward updating rule, as specified in Refs. \cite{huepe.c:2008,baglietto.g:2009}.
In most calculations presented here, we assume a constant propulsion speed $v_0=1$ and time step $\tau=1$. In some runs we use smaller
particle velocities $v_0=0.05$ and $v_0=0.1$ to test the stability of the theory predictions. In all simulations we keep the density
$\rho_0=3.18$ and the radius of the interaction $R=0.1$ constant.
This corresponds to a fixed small average number of collision partners, $M=\pi R^2 \rho_0=0.1$. We did not use larger $M$, mainly because
the kinetic theory is easier to apply in the binary collision approximation which is valid for $M\ll 1$.

\subsection{Motion statistics}

To analyze the collective behavior of the model we perform a series of simulations changing  the noise strength, $\eta$, and interaction angle $\alpha$. We characterise the orientational ordering by the polar order parameter \cite{vicsek.t:1995,chate.h:2008,baglietto.g:2009,nagy.m:2007,romenskyy.m:2013}
\begin{equation}
\varphi = \left \langle  \frac{1}{N} \left |\sum_{i = 1 }^N \exp(\i \theta_i) \right | \right \rangle
\label{order_param}
\end{equation}
where $\i$ is the imaginary unit and $\theta_i$ is the direction of motion of particle $i$. This order parameter turns zero in the isotropic phase and assumes finite positive values in the ordered phase, which makes it easy to detect the transition. However, at low densities it may be more difficult to detect the transition due to the relatively small number of particles constituting the system and large density fluctuations. To locate transition points precisely we also calculated the Binder cumulant \cite{binder.k:1981}
\begin{equation}
G_L = 1- \frac{\langle \varphi^4_L \rangle_t}{3\langle \varphi^2_L \rangle^2_t}
\label{binder}
\end{equation}
where $\langle \cdot \rangle_t$ stands for the time average and index $L$ denotes the value calculated in a system of size $L$. The most important property of the Binder cumulant is a very weak dependence on the system size so $G_L$ takes a universal value at the critical point, which can be found as the intersection of all the curves $G_L$ obtained at different system sizes \cite{chate.h2:2008} at fixed density. To detect the transition points in $\eta-\alpha$ plane precisely we plot three curves for different $L$ at constant density and find the point where they cross each other. Then, we use those points to construct the phase diagram.

To characterise the apolar ordering within our models in unconfined space we use the following nematic order parameter
\begin{equation}
Q=\left \langle \left|\frac{1}{N}\sum_{i = 1 }^N \exp(\i2\theta_i)\right| \right \rangle.
\label{apolar_param}
\end{equation}
When the motion of particles is perfectly collinear, irrespective of the direction of motion, $Q$ equals 1. We note that a perfectly polarly ordered phase is characterized by $\varphi =Q = 1$, as the polar ordering implies the nematic ordering. An apolarly ordered phase requires only $Q = 1$ while the polar order parameter can take any value $\varphi < 1$. Therefore, requirements for polar order are more restrictive.

\section{Results}
\label{results}
\begin{figure}
\centering
\subfigure[]{
\includegraphics[width=6.5cm,clip]{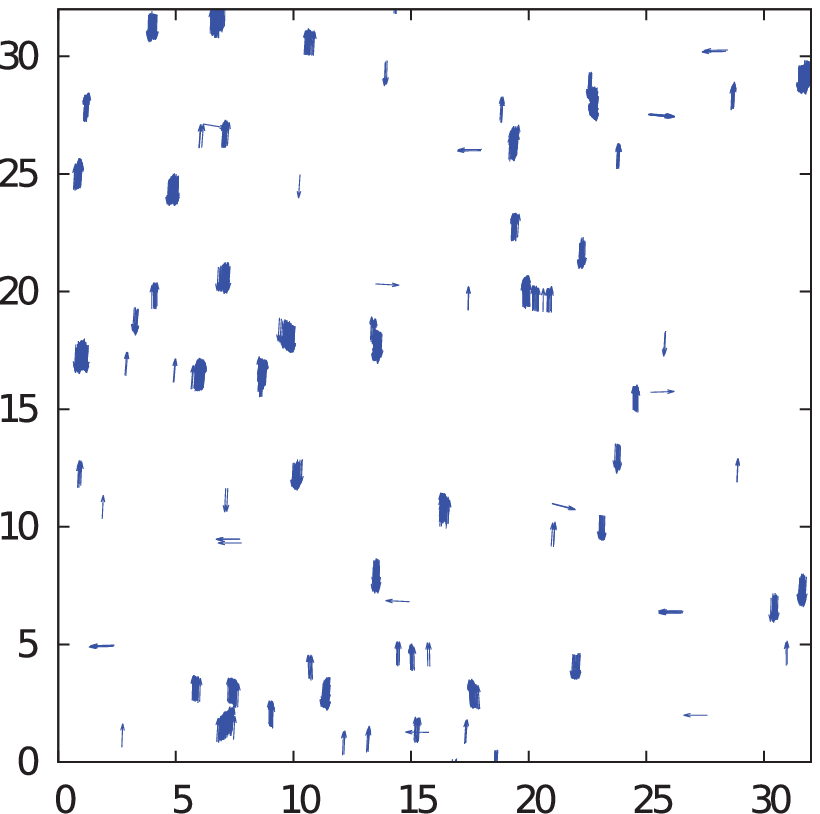}
}
\subfigure[]{
\includegraphics[width=6.5cm,clip]{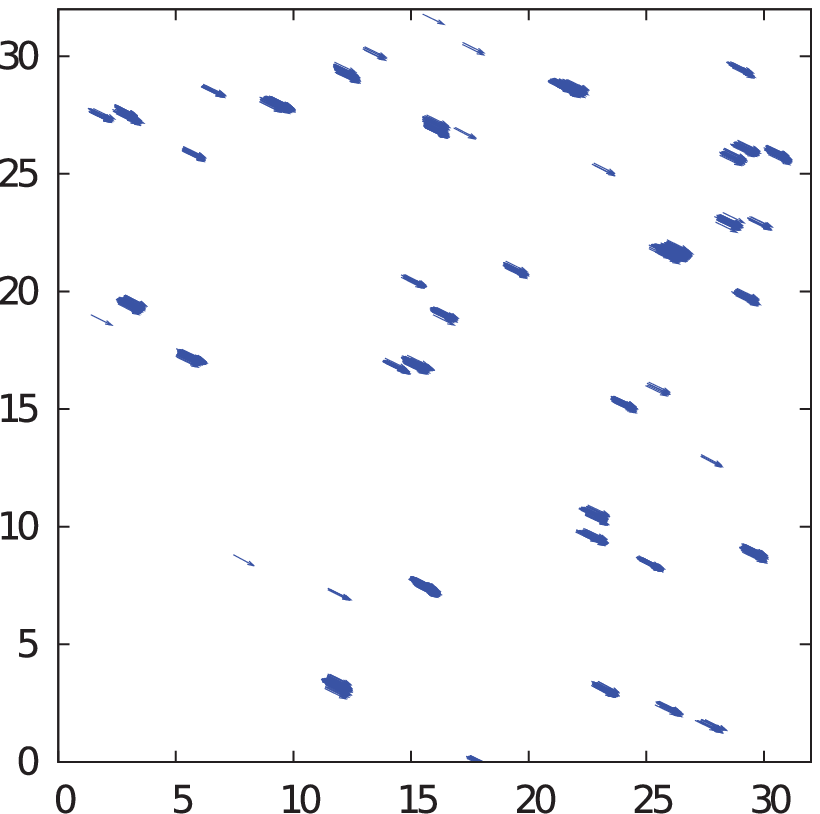}
}
\caption{Typical steady-state snapshots of the restricted angle Vicsek model at different values of the restriction angle $\alpha$ ($v_0=1$, $L=32$, $\eta=0$). (a) $\alpha=0.35\pi$. (b) $\alpha=1\pi$.}
\label{fig:vra_snapshot1}
\end{figure}

First, we look at the collective behavior of the system upon variation of the restriction angle $\alpha$ in absence of noise (Figs.~\ref{fig:vra_snapshot1}). When $\alpha$ is small [Figs.~\ref{fig:vra_snapshot1}(a)], $\alpha=0.35 \pi$, we see many well-packed groups of particles moving collinearly albeit often in opposite directions. At the same time, one can observe significant number of single particles or small clusters consisting of two-three particles that move almost perpendicular to large ones. At $\alpha=\pi$ [Fig.~\ref{fig:vra_snapshot1}(b)], larger clusters moving in the same direction are formed.

\begin{figure}
\centering
\subfigure[]{
\includegraphics[width=6.5cm,clip]{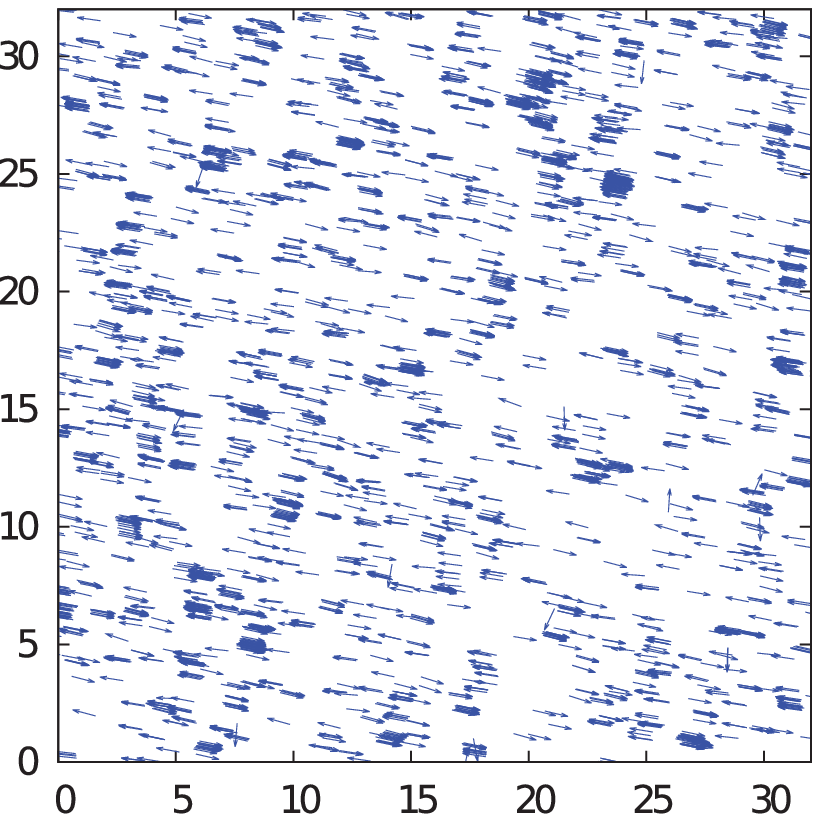}
}
\subfigure[]{
\includegraphics[width=6.5cm,clip]{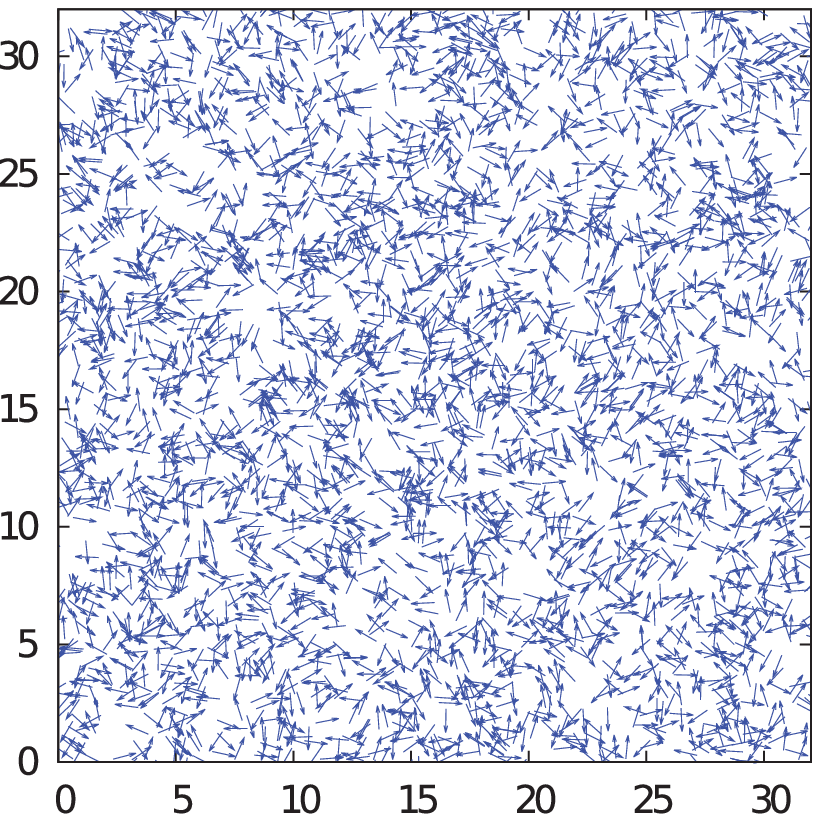}
}
\caption{Typical simulation snapshots of the restricted angle Vicsek model at different noise values ($v_0=1$, $L=32$, $\alpha=0.35 \pi$). (a) $\eta=0.125$. (b) $\eta=0.29$.}
\label{fig:vra_snapshot2}
\end{figure}

Next, we look at the behavior of the system under variation of the noise amplitude $\eta$ at a fixed restriction angle $\alpha=0.35 \pi$. We have previously seen on Fig.~\ref{fig:vra_snapshot1}(a) that at zero noise large clusters are formed, which move collinearly in opposite directions. Figs.~\ref{fig:vra_snapshot2}(a) and (b) show snapshots for two more noise values $\eta=0.125$ and $0.29$. At $\eta=0.125$, we observe more small oppositely aligned clusters than at $\eta=0$. At the higher noise level, $\eta=0.29$, no clustering occurs and velocities of all particles are distributed randomly.

\begin{figure}
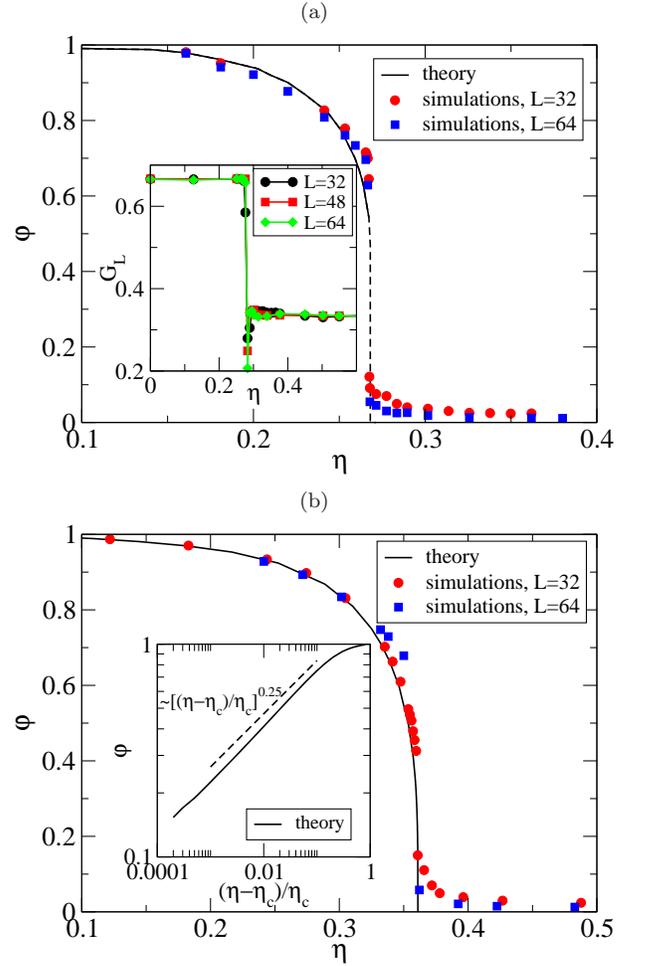

\centering
\subfigure[]{
\includegraphics[width=8cm,clip]{a035pi_with_theory.eps}
}
\subfigure[]{
\includegraphics[width=8cm,clip]{a0443pi_with_theory.eps}
}
\caption{Polar order parameter $\varphi$ plotted against noise $\eta$ for the zone of the first order phase transition ($v_0=1$, theoretical curves are obtained by the algorithm with high $k$ inclusion, see section \ref{sec:fredholm}). (a) $\alpha=0.35\pi$. \emph{Inset:} Binder cumulant for three different system sizes. (b) $\alpha=0.443\pi$. \emph{Inset:} Phase diagram for $\alpha=0.443\pi$. Here, $k_{max}=500$ was chosen for the theoretical calculations.}

\label{fig:vra_polar_lowalpha}
\end{figure}

\begin{figure}
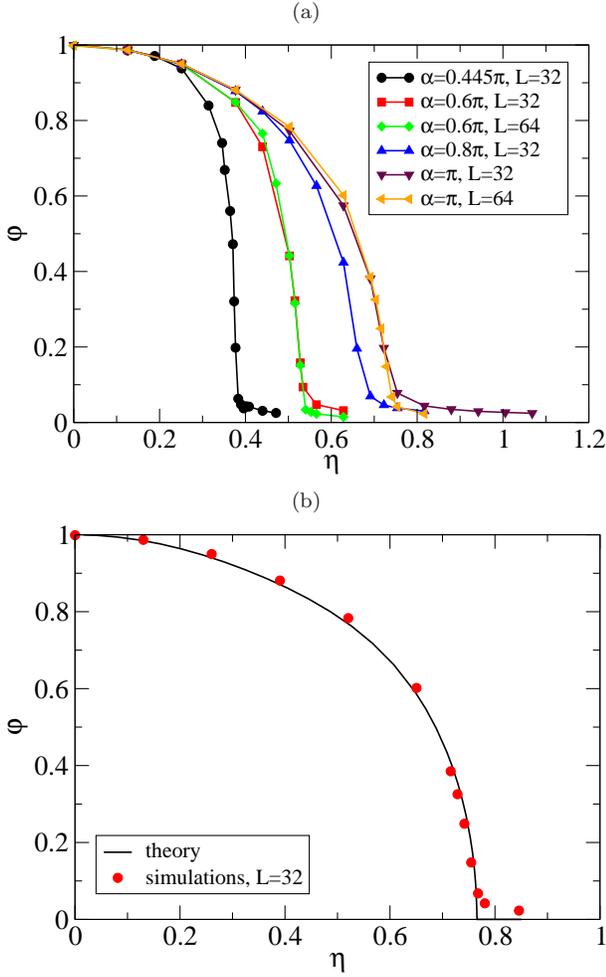

\subfigure[]{
\includegraphics[width=8cm,clip]{vra_order_second.eps}
}
\subfigure[]{
\includegraphics[width=8cm,clip]{a1pi_with_theory.eps}
}
\caption{Polar order parameter $\varphi$ plotted against noise $\eta$ for the zone of the second order phase transition ($v_0=1$). (a) $\varphi$ for different system sizes (the lines are a guide to the eye). (b) Comparison of the theoretical and computational results for $\alpha=\pi$.}
\label{fig:vra_polar_largealpha}
\end{figure}

In Figs.~\ref{fig:vra_polar_lowalpha} and \ref{fig:vra_polar_largealpha} we plotted the iso-$\rho$ curves for different restriction angles $\alpha$ and system sizes $L$. For $\alpha=0.35\pi$ we observe a clearly defined first order phase transition at all system sizes [Fig.~\ref{fig:vra_polar_lowalpha}(a)]. At a higher value of $\alpha$ ($\alpha=0.443\pi$), the first order phase transition becomes less pronounced [Fig.~\ref{fig:vra_polar_lowalpha}(b)], however for systems with linear size $L=48$ and $64$ values of the order parameter close to the transition point are much higher than those for the smaller system $L=32$. Finally, for all values of $\alpha$ higher that $0.443\pi$ the transition seems to change into a continuous one (Fig.~\ref{fig:vra_polar_largealpha}). Thus, we have an apparent tricritical point at $\alpha\approx0.444\pi$.

\begin{figure}
\centering
\includegraphics[width=8cm,clip]{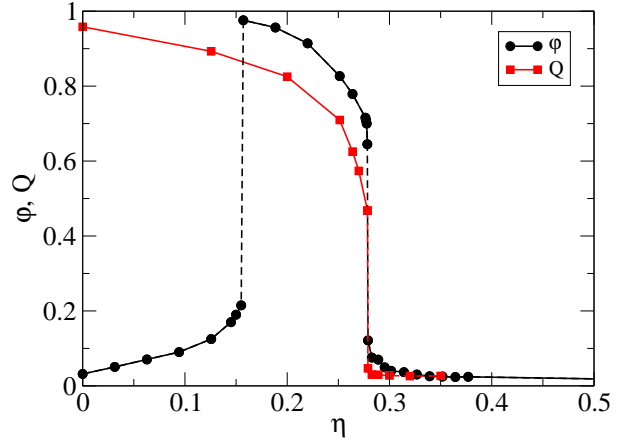}
\caption{Polar $\varphi$ and nematic order parameter $Q$ plotted against noise $\eta$ for $L=32$, $\alpha=0.35 \pi$ and $v_0=1$ (the lines are a guide to the eye).}
\label{fig:vra_apolar}
\end{figure}

\begin{figure}
\centering
\includegraphics[width=8cm,clip]{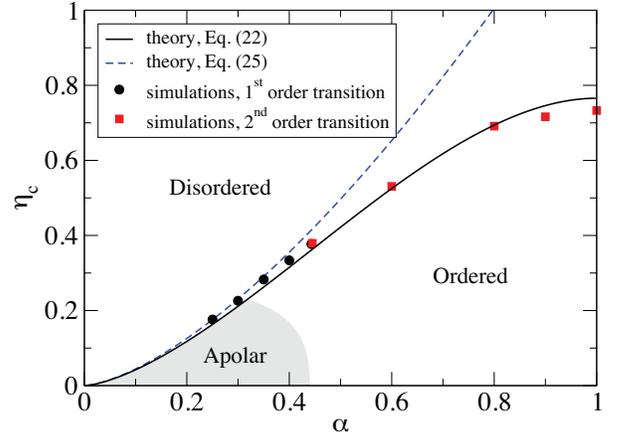}
\caption{Phase diagram for the restricted angle Vicsek model in a square simulation box ($v_0=1$). Solid line shows the prediction of Eq. \eqref{GAMMA_EQ} for $M=0.1$. $\alpha$ is given in units of $\pi$. Shaded area shows the region of stability of the apolar phase in simulations.}
\label{fig:vra_phasediag}
\end{figure}

We should also note that at low restriction angles our system exhibits strong apolar alignment (see Fig.~\ref{fig:vra_apolar}) with the values of the nematic order parameter $Q \approx 1$ at zero noise. This finding is in agreement with the observed distribution of particle orientations shown in Figs.~\ref{fig:vra_snapshot1}(a),(b). At $\alpha=0.35 \pi$ with zero noise, the nematic order parameter is close to unity as the majority of clusters move collinearly in opposite directions. The apolar ordered phase disappears at $\eta \approx 0.29$. The appearance of the apolar state is a result of ``polarization'' which is caused by the limited angle of view in the velocity plane and is analyzed in
chapter \ref{sec:nematic}. Thus, if $\alpha$ is small and two groups of particles meet coming from different directions, the members of each group can use only the neighbors in their own pack for orienting themselves, and the clusters continue moving in the same direction as before the ``collision''. Once the ordered apolar state develops it is very stable: during the whole simulation time ($10^7$ time steps) the state of the system does not change and the stable apolar ordering cannot be avoided by changing the initial conditions.
We have checked the stability of the apolar steady state by periodically ``shaking'' the system by short periods of stronger noise. We found that the value of the polar order parameter $\varphi$ does not return to the previous value after the shake while the value of the apolar order parameter $Q$ is always recovered even if we start from a configuration with a perfect polar order. We therefore conclude that the apolar ordered state is more stable in the region indicated in Fig~\ref{fig:vra_phasediag}.

\begin{figure}
\centering
\subfigure[]{
\includegraphics[width=4.0cm,clip]{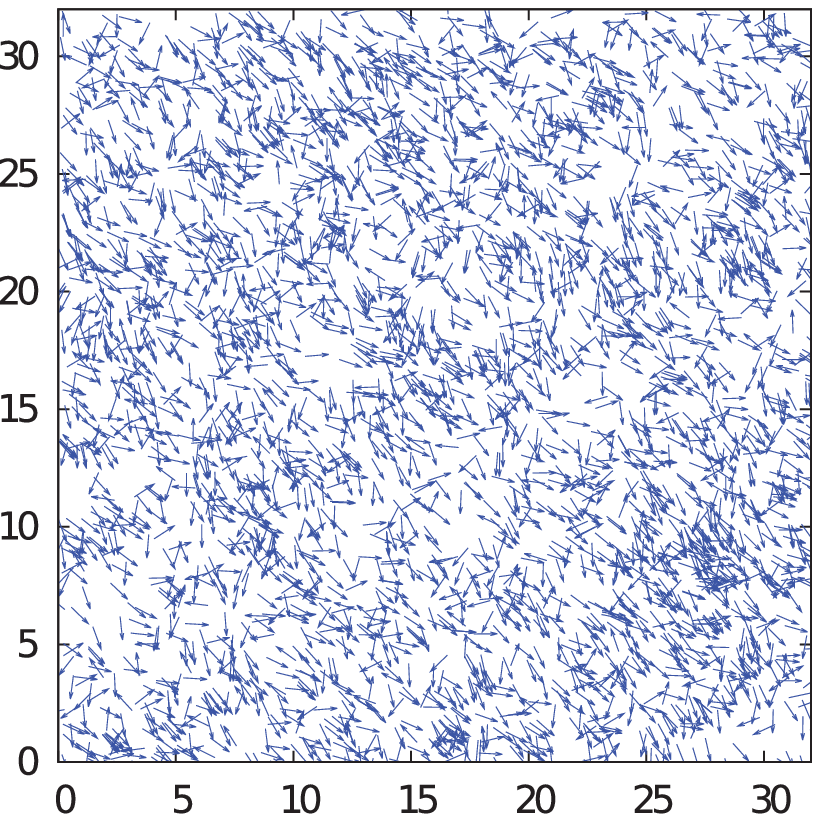}
}
\subfigure[]{
\includegraphics[width=4.0cm,clip]{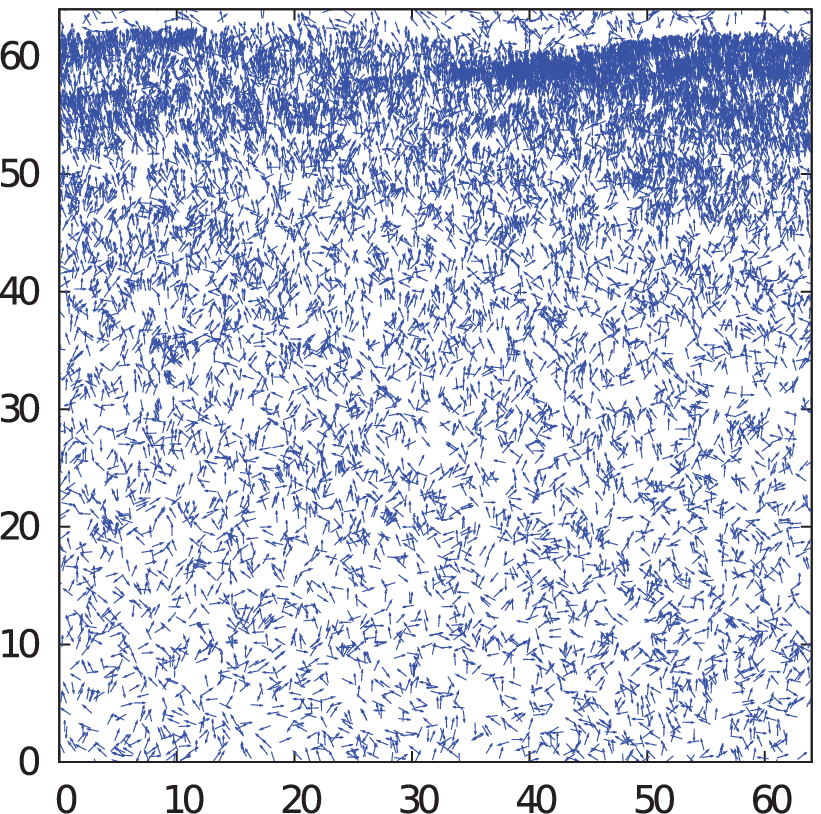}
}
\caption{Typical simulation snapshots of the restricted angle Vicsek model at different system sizes ($v_0=1$, $\eta=0.35$, $\alpha=0.443$). (a) $L=32$. (b) $L=64$.}
\label{fig:vra_snapshot_L}
\end{figure}

It has been argued previously \cite{chate.h:2008,ihle.t:2013} that the linear system size has a significant influence on critical behavior in the VM.
In particular, it may lead to a change in a cluster distribution and formation of traveling waves, which may in
its turn also change the kind of the phase transition. We already observed an unusual behavior of the order parameter upon variation of the system size [Fig. \ref{fig:vra_polar_lowalpha}(b)]. To study this question in more detail we compare the snapshots for two different box lengths $L$ as obtained for a noise strength close to the critical value (Fig. \ref{fig:vra_snapshot_L}). For $L=32$ we observe a quite homogeneous distribution of the particles across the simulation box while for $L=64$ we see a large density wave traveling in the $y$-direction. A similar density band is observed for $L=48$. Based on these observations, we can speculate that density bands are responsible for the rise of the order parameter for two system sizes, $L=48$ and $64$, that we have seen in Fig. \ref{fig:vra_polar_lowalpha}(b).
The use of the term ``phase transition'' for systems of finite size is discussed in Appendix B.

The state diagram for our system is shown in Fig.~\ref{fig:vra_phasediag}. At small restriction angles $\alpha$, the transition to the orientationally ordered phase happens only at very small noise amplitudes, which can be explained by the high level of polarization of the individual clusters.
We also see that the critical noise amplitude seems to be proportional to the interaction angle at the transition
line $\eta_c \propto \alpha$ in the measured low-$\alpha$ region, $0.25\le\alpha\le 0.6$.
However, as discussed earlier, this apparent linear behavior is still consistent with the asymptotic power-law behavior, $\eta_c\sim\alpha^{3/2}$ which is  theoretically predicted for $\alpha\rightarrow 0$. This power law is given by the dashed curve in
Fig. \ref{fig:vra_phasediag}.
It is clear from this plot that the exponent $3/2$ as opposed to $1$ could only be detected
at $\alpha<0.2$, that is, at values too small to be probed by the agent-based simulations.
Nevertheless, the data points from the simulations agree very well with the curve given by the full theory, Eq. (\ref{GAMMA_EQ}).

\begin{figure}
\centering
\subfigure[]{
\includegraphics[width=8cm,clip]{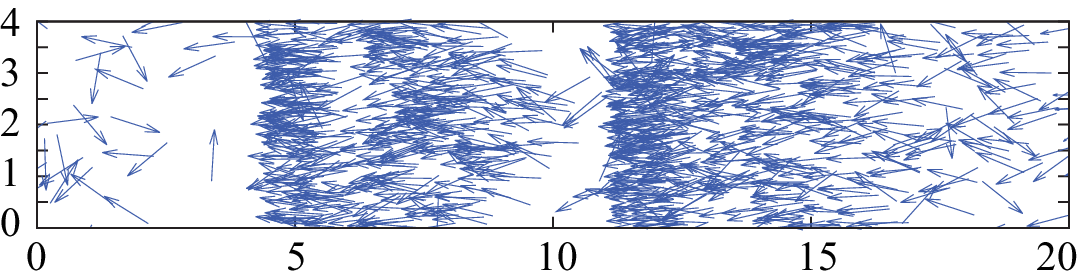}
}
\subfigure[]{
\includegraphics[width=8cm,clip]{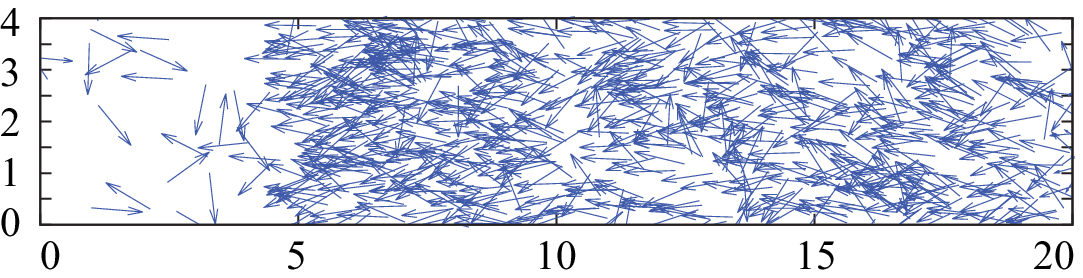}
}
\caption{Typical steady-state simulation snapshots taken close to transition point ($v_0=1$, $L_x=128$, $L_y=4$, only part of the simulation box is shown for clarity). (a) $\alpha=0.443\pi$, $\eta=0.34$. (b) $\alpha=\pi$, $\eta=0.75$.}
\label{fig:vra_snapshot_e}
\end{figure}

As density bands are often considered to be a signature of the first order phase transition, it is interesting to
see what happens if we have those waves in a system with values of $\alpha$ close or equal to $0.444\pi$, above which
a continuous transition is expected in small systems.
To enhance formation of the density waves we have run series of simulations in very elongated boxes: $L_x=128$, $256$ and $512$. Another dimension, $L_y$ in all three cases was kept constant ($L_y=4$). Two snapshots for a system of a linear size $128\times4$ are shown in Fig. \ref{fig:vra_snapshot_e}. At $\alpha=0.443\pi$ we see two compact waves traveling along the $x$-axis. For $\alpha=\pi$ we observe one large wave also moving in $x$-direction.

\begin{figure}
\centering
\includegraphics[width=8cm,clip]{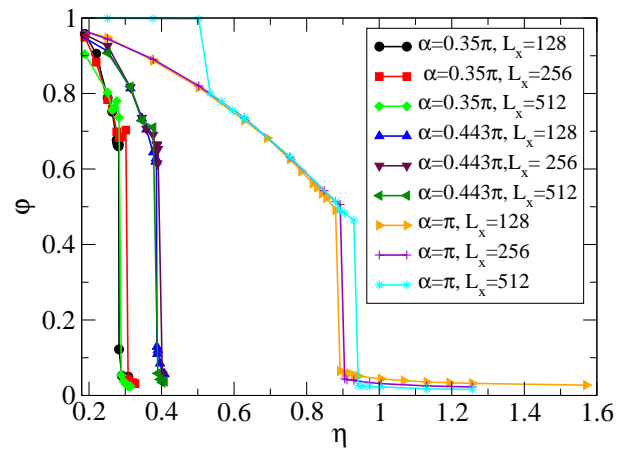}
\caption{Polar order parameter $\varphi$ plotted against noise $\eta$ for various values of $\alpha$ and $L_x$ ($v_0=1$, simulations with elongated box, the lines are a guide to the eye).}
\label{fig:vra_polar_e}
\end{figure}

In Fig. \ref{fig:vra_polar_e}, we show a plot of the orientational order parameter for simulations in the elongated box. For all values of $\alpha$ as well as for all system sizes, we observe a sharp drop of the order parameter at the transition point, indicating a discontinuous phase transition (also confirmed by the Binder cumulant analysis). For the larger system sizes, the transition to the disordered state happens at the higher noise levels. This behavior can be explained by the formation of larger density waves \cite{ihle.t:2013} and is in agreement with the observations made earlier for the square box.

\begin{figure}
\centering
\includegraphics[width=8cm,clip]{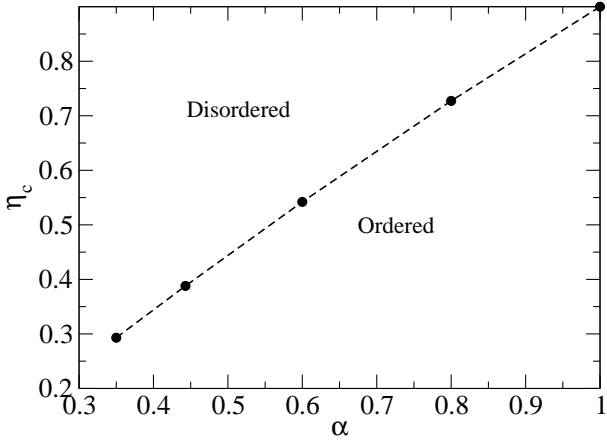}
\caption{Phase diagram for the restricted angle Vicsek model with elongated simulation box ($v_0=1$). $\alpha$ is given in units of $\pi$ (the line is a guide to the eye).}
\label{fig:vra_phasediag_e}
\end{figure}

An example of the phase diagram for a system in an elongated box is shown in Fig. \ref{fig:vra_phasediag_e}. If we compare it with the phase diagram for the square box, we indeed see three clearly distinguishable differences: the transition in the elongated box happens at a higher noise level while the dependence of $\eta$ on $\alpha$ appears to be linear for the entire range of $\alpha$ values used.
Note, that this does not mean that the asymptotic dependence for $\alpha\rightarrow 0$ has to be linear. For example, for small boxes,
kinetic theory predicts $\eta_c\sim\alpha^{3/2}$, see Eq. (\ref{SMALL_ALPHA}).
Furthermore, the transition is now discontinuous for all $\alpha$, that is, the tricritical point has disappeared due to the formation of density waves that made the transition discontinuous even at larger $\alpha$.
This type of ``finite-size effect'' is qualitatively different from the ones observed in the (grand)canonical ensemble of
equilibrium systems such as the Potts-model.

\begin{figure}
\centering
\includegraphics[width=8cm,clip]{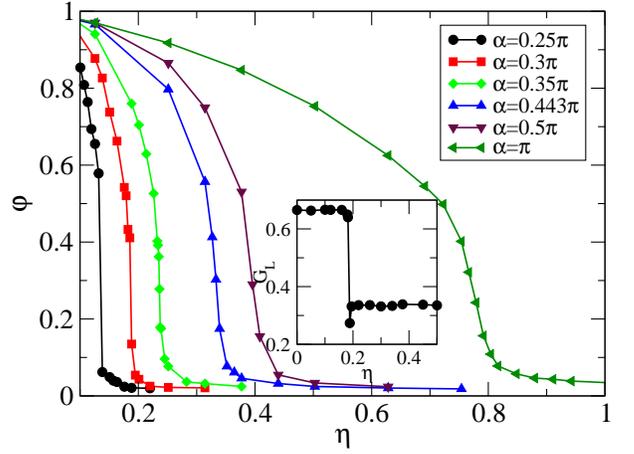}
\caption{Polar order parameter $\varphi$ plotted against noise $\eta$ for various values of $\alpha$ ($v_0=0.1$, $\lambda/R=1$, the lines are a guide to the eye). \emph{Inset:} Binder cumulant $G_L$ plotted against noise $\eta$ for $\alpha=0.3 \pi$ ($v_0=0.1$, $L=32$).}
\label{fig:vra_polar_v01}
\end{figure}

\begin{figure}
\centering
\includegraphics[width=8cm,clip]{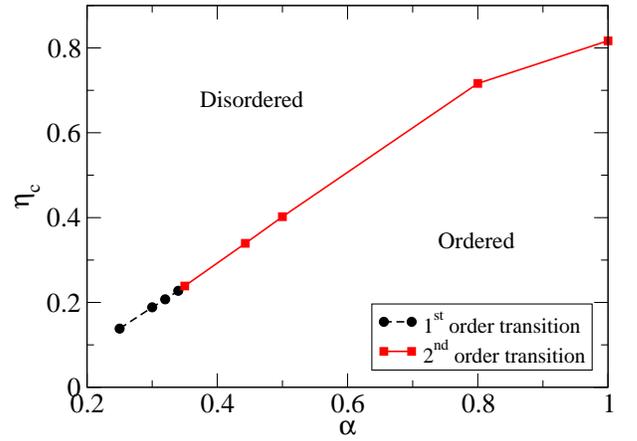}
\caption{Phase diagram for the restricted angle Vicsek model at $v_0=0.1$ ($\lambda/R=1$). $\alpha$ is given in units of $\pi$, the line is a guide to the eye.}
\label{fig:vra_phasediag_e1}
\end{figure}

\begin{figure}
\centering
\includegraphics[width=8cm,clip]{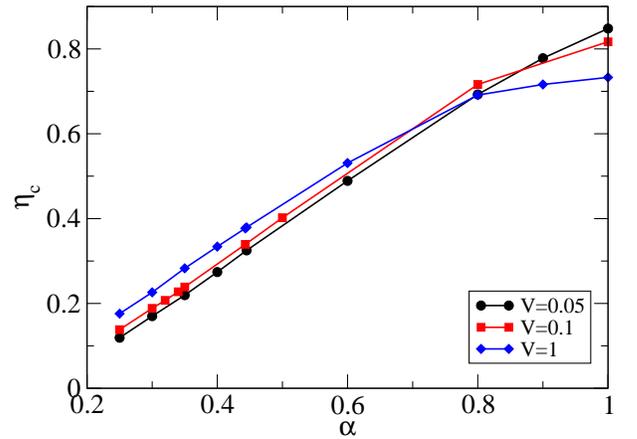}
\caption{Phase diagrams for the restricted angle Vicsek model at different particle speeds $v_0$: $v_0=0.05$ ($\lambda/R=0.5$), $v_0=0.1$ ($\lambda/R=1$), and $v_0=1$ ($\lambda/R=10$). $\alpha$ is given in units of $\pi$, the lines are a guide to the eye.}
\label{fig:vra_phasediag_e2}
\end{figure}

In Figures \ref{fig:vra_polar_v01}, \ref{fig:vra_phasediag_e1} and \ref{fig:vra_phasediag_e2} we study the influence of the particle speed.
All of our previous simulations were performed at a very large ratio of the mean free path $\lambda=v_0\tau$ to the interaction radius $R$,
$\lambda/R=10$. In this high speed regime, kinetic mean-field theory is supposed to be very accurate.
A change of the phase behavior is expected if this ratio
is reduced to below unity.
This is because in the low velocity regime, $\lambda/R\ll 1$, pre-collisional correlations are expected to be strong and the Molecular Chaos approximation is supposed to fail, see discussion in Ref. \cite{ihle.t.1:2014}.
Here, we dropped $\lambda/R$ from 10 to 1 and 0.5 and only observe moderate changes of the threshold value $\eta_c$, see Fig.
 \ref{fig:vra_phasediag_e2}. We hypothesize, that $\lambda/R=0.5$ is still too large to see a significant reduction of $\eta_c$.
What is more interesting is that the tricritical point drops down from $0.443\pi$ to about $0.35\pi$ at $v_0=0.05$, see Fig.
\ref{fig:vra_phasediag_e1}.
It has been shown in Ref. \cite{ihle.t.1:2014} that in the regular VM
the average number of interaction neighbors increases due to clustering when the ratio $\lambda/R$ is lowered.
It is natural to assume that similar clustering occurs in the bounded-confidence model.
A possible explanation of the drop of $\alpha_c$ could then be that the angular restriction mechanism is less relevant due
to the increased supply of neighbors. In other words, if there are many neighbors to choose from, it is not as dramatic to exclude, for example, two thirds of them as there is still somebody left with the ``right mind-set'' to interact with.
Alternatively,
the reduction of $\alpha_c$ could also just be a consequence of the reduced threshold noise in the $\alpha$-range around $0.4\pi$, as seen in Fig. \ref{fig:vra_phasediag_e2}.

\section{Discussion}
\label{discussion}
Introducing selectivity into the Vicsek model leads to a new qualitative feature: Within homogeneous mean-field theory,
the transition from disorder to homogeneous states of
collective motion can be either continuous or discontinuous, depending on the restriction parameter $\alpha$.
As a consequence, tricritical points, which separate regions of different transition behavior, appear.
In a social context, the possibility of discontinuous transitions
already in small systems, and the
occurrence
of tricritical points,
can be rationalized as follows:
The noise amplitude $\eta$ describes the degree of how particles conform to what the neighbors they are ``listening to''
do. Small $\eta$ means a high degree of conformity. If agents are highly ignorant of opposing views, that is if $\alpha$ is small,
change to collective behavior from disordered motion and back can only happen in an abrupt, discontinuous fashion.
Let us assume particles are initially ordered and the noise is increased. That means the agents deviate more and more
from what their tolerated neighbors suggest. Since the agents are in an ordered state, most of their neighbors are already aligned with
their common ``opinion'',  and they only have to disregard a few neighbors, even though $\alpha$ is small. In other words, the angular restriction
is not that important in a highly ordered state. If the noise $\eta$ is slightly increased, this situation can persist because just a few more
neighbors have to be ignored but there is still a sufficient amount of like-minded ones in reach.
By filtering out the opinions even if they only slightly deviate from their own, the agents can maintain their collective
motion even when the noise is slightly above $\eta_c$.
However, once the noise is too high, the agents can barely find anyone anymore who fits their narrow focus and with whom they can align with.
As a result, everyone disregards almost anyone else, particles just perform interaction-free random walks,
and the global order collapses abruptly.
Thus, in this disordered state, the angular restriction has a large impact as compared to within the ordered state.
That means, going back to the ordered state starting from disorder by decreasing the noise will require a substantial decrease of
the noise to a value well below $\eta_c$.
Once the restriction angle is larger than $\alpha_c$ the selection mechanism is not effective anymore and the system
qualitatively behaves like the standard VM. In this case, a phase transition is observed that is continuous
at the mean-field level if one suppresses the formation of
inhomogeneous stationary states which consist of soliton-like invasion waves, Refs. \cite{gregoire:2004,chate.h:2008,aldana.m:2009}.
This is because using inhomogeneous mean-field theory these waves have been shown to render
the phase transition discontinuous, Ref. \cite{ihle.t:2013}.
Since these waves grow from a long-wavelength instability and have a minimum spatial extent, one just has to keep the system size below a critical
size $L^*$, to prevent their occurrence. Once the system size exceeds $L^*$, the order/disorder transition is discontinuous
(at least in the large speed regime considered here), even above $\alpha_c$. However, the mechanism for the discontinuity is now
very different from the one acting at low $\alpha$ and in small systems
(for a general discussion on transitions in small systems, see Appendix B.)

The existence of a discontinuous transition in small systems without the invasion-wave mechanism can also be rationalized by
comparing it to the VM with vectorial noise, see Refs. {\cite{gregoire:2004,aldana.m:2007}:
In this model, a vector of fixed length but with random orientation is added to the vector sum $\mathbf{A}$
of all pre-collisional velocities in the collision circle.
This addition has a large impact on the final orientation if the size of $\mathbf{A}$ is small, that is, if the system is in the disordered state, but
is less relevant in the ordered state where $\mathbf{A}$ is large. Thus, as in the current model, the influence of noise depends on
whether the pre-collisional state is ordered or disordered.
It is this dependence on initial conditions on the microscopic level that allows for hysteresis -- an indicator
for a discontinuous phase transition. In the regular VM with angular noise, the impact of noise does not depend on whether
$\mathbf{A}$ is long or short -- in both cases $\mathbf{A}$ is rotated by the same amount.
Thus, in the regular VM, hysteresis is only possible on a mesoscopic level, through the formation of strongly inhomogeneous band-like states.

\section{Conclusions}
\label{summary}
We introduced a version of the Vicsek model of self-propelled particles with selective interactions, such that the particles can align only with the neighbors whose direction of motion is not too different from their own. We developed a mean-field theory of the flocking dynamics in such a system and predicted its dynamic phase diagram. In particular, we find that
in homogeneous systems,
depending on the interaction restriction angle and the system size, the transition between the ordered and disordered states of the system can be continuous or discontinuous. The discontinuous behavior is observed in small systems at small interaction angles. We predict the position of the transition line and of the tricritical point and suggested an interpretation of the results in terms of opinion dynamics, where the restriction angle reflects the bounded confidence. We tested the theory using direct simulations of self-propelled particles and found excellent agreement with the kinetic theory predictions. We also observed that at very small interaction angles the polar ordered phase becomes unstable with respect to the apolar
phase.

We discovered that, below a certain restriction angle, apolar ordered solutions of the kinetic equations exist
such as nematic and higher order fragmented states.
We calculated the critical noise, at which the disordered state bifurcates to a nematic state.
This calculation has been generalized to systems that show fragmentation into more than two groups. A scaling law for
the transition to collective motion for an arbitrary number of distinctly oriented groups has been obtained.
Our kinetic theory predicts two different tricritical points for transitions from disorder to a state with at least two distinct subgroups:
At low and high restriction angle the transition is discontinuous but continuous at intermediate
$\alpha$.
The maximum restriction angle, below which fragmented states can exist, is found to be inversely proportional to the number of subgroups.
This finding is consistent with studies on social network models with bounded confidence \cite{kurmyshev.e:2011}:
A smaller opinion tolerance (corresponding to a smaller restriction angle $\alpha$) allows for a stronger fragmentation of the opinion.
More research is needed to investigate the stability and competition of the fragmented states.
Note that the mean-field theory used in this paper reaches its limits for large numbers of subgroups:
The noise amplitudes required to observe these states are so small that the underlying molecular chaos assumption
is violated, even at quite large time steps and particle velocities.
This makes it difficult to achieve quantitative agreement between kinetic theory and agent-based simulations of fragmented states.
To properly describe these states analytically, an approach beyond mean-field, such as the ring-kinetic theory of Ref. \cite{chou.yl:2014}
is required. In general, the selective interactions introduced in this paper could make models of social agents more realistic.
Furthermore, the proposed kinetic theory could be helpful for a closely related version of the Vicsek model, which was introduced by
Lu {\em et al.} \cite{lu.s:2013} to describe their experiments on the collective motion of {\em Bacillus subtilis}.

Finally, we presented a novel algorithm to accurately solve a nonlinear Fredholm equation with a singular kernel that occurs in the kinetic
theory of Vicsek-style models. This algorithm delivers highly accurate results for the one-particle distribution function
as a function of noise, not only close to the order/disorder threshold but for all possible noises,
even down to almost zero noise, where the order parameter approaches unity.

\begin{acknowledgments} Support from the National Science Foundation under grant No. DMR-0706017 (TI), Irish Research Council for Science, Engineering and Technology, Grant RS/2009/1788 (MR and VL), and European Research Council grant IDCAB 220/104702003 (MR) is gratefully acknowledged.
TI would like to thank C. Huepe for valuable discussions.
\end{acknowledgments}

\appendix
\section{Calculation of the coupling integrals}

The formula for the coupling coefficients $B^{(2)}_{kpq}$, Eq. (\ref{COUPLING_B2}), can in principle
be fed into a symbolic algebra software like Mathematica.
However, it is still useful to discuss these coefficients in more detail and to give specific examples
which are needed to derive the threshold noise and the phase diagram of the restricted-angle Vicsek model.
First, Eq. (\ref{COUPLING_B2}) is rewritten as
\begin{eqnarray}
\label{APP_EQ1}
B^{(2)}_{kpq}(\alpha)&=&{1\over (2\pi)^2} \int_0^{2\pi}\,d\theta_1{\rm cos}(p\theta_1)
\,U_{kq}\;\;\;{\rm with} \\
\label{APP_EQ2}
U_{kq}(\alpha)&=&\int_{\theta_1-\alpha}^{\theta_1+\alpha}{\rm cos}[k(\theta_1+\theta_2)/2]
\,{\rm cos}(q\theta_2)\,d\theta_2\:\:\:
\end{eqnarray}
Here and in the following, we assume a non-negative restriction angle, $0\leq \alpha \leq \pi$.
Using trigonometric identities, the auxiliary function $U_{kq}(\theta_1)$, can be expressed as
\begin{equation}
\label{APP_EQ3}
U_{kq}={\rm cos}\left({k\over 2}\theta_1\right)P_{kq}
-{\rm sin}\left({k\over 2}\theta_1\right)Q_{kq}
\end{equation}
with another set of auxiliary functions $P_{kq}$ and $Q_{kq}$.
Integrating over $\theta_2$ gives,
\begin{eqnarray}
\nonumber
& &P_{kq}=\alpha[\delta_{q,k/2}+\delta_{q,-k/2}] \\
\nonumber
&+&{[1-\delta_{q,k/2}] \over 2\left(q-{k\over 2}\right)}
\left\{ {\rm sin}\left[\left(q-{k\over 2}\right)\left(\theta_1+\alpha\right)\right] \right. \\
\nonumber
& & -\left. {\rm sin}\left[\left(q-{k\over 2}\right)\left(\theta_1-\alpha\right)\right]
\right\} \\
\nonumber
&+&{[1-\delta_{q,-k/2}] \over 2\left(q+{k\over 2}\right)}
\left\{ {\rm sin}\left[\left(q+{k\over 2}\right)\left(\theta_1+\alpha\right)\right] \right. \\
\label{APP_EQ4}
& &-\left. {\rm sin}\left[\left(q+{k\over 2}\right)\left(\theta_1-\alpha\right)\right]
\right\}
\end{eqnarray}
\begin{eqnarray}
\nonumber
& &Q_{kq}=
{[1-\delta_{q,k/2}] \over 2\left(q-{k\over 2}\right)}
\left\{ {\rm cos}\left[\left(q-{k\over 2}\right)\left(\theta_1+\alpha\right)\right] \right. \\
\nonumber
& & -\left. {\rm cos}\left[\left(q-{k\over 2}\right)\left(\theta_1-\alpha\right)\right]
\right\} \\
\nonumber
&-&{[1-\delta_{q,-k/2}] \over 2\left(q+{k\over 2}\right)}
\left\{ {\rm cos}\left[\left(q+{k\over 2}\right)\left(\theta_1+\alpha\right)\right] \right. \\
\label{APP_EQ5}
& &-\left. {\rm cos}\left[\left(q+{k\over 2}\right)\left(\theta_1-\alpha\right)\right]
\right\}
\end{eqnarray}
After substituting these expressions back into Eqs. (\ref{APP_EQ3}) and (\ref{APP_EQ1})
the final integration over $\theta_1$ can be easily performed.
Instead of giving this lengthy expression in full generality, we will only discuss a few relevant cases.
For example, for $k=1$ and $q=0,1$ one finds
\begin{eqnarray}
\nonumber
U_{10}&=&4 {\rm sin}\left({\alpha\over 2}\right)\,{\rm cos}(\theta_1) \\
U_{11}&=&2 {\rm sin}\left({\alpha\over 2}\right)+{2\over 3} {\rm sin}\left({3\alpha\over 2}\right)\,
{\rm cos}(2\theta_1)
\end{eqnarray}
Inserting this into Eq. (\ref{APP_EQ1}) gives,
\begin{equation}
\label{B2_110_RESULT}
B^{(2)}_{101}=
B^{(2)}_{110}={1\over \pi}  {\rm sin}\left({\alpha\over 2}\right)
\end{equation}
Thus, as expected, $B^{(2)}_{kpq}$ is symmetric in the indices $p$ and $q$.
However, $B^{(1)}_{kpq}$ does not have this symmetry. For example,
from Eq. (\ref{COUPLING_B1}) one obtains,
\begin{eqnarray}
\nonumber
B^{(1)}_{110}&=&{1\over 2}\left(1-{\alpha \over \pi}\right) \\
B^{(1)}_{101}&=&- { \sin{\alpha} \over 2 \pi}
\end{eqnarray}
Hence, the total coupling constant $B_{kpq}=B^{(1)}_{kpq}+B^{(2)}_{kpq}$ is not symmetric, $B_{kpq}\neq B_{kqp}$.
The physical reason for this is that the focal particle (with corresponding Fourier index $p$) plays a special role in the collision:
If the angular difference between particles 1 and 2 is too large, it is always particle 1 that is allowed to determine
the mean direction and never particle 2.
In other words, the social bias of an agent to favor its own ``opinion'' leads to asymmetric coupling matrices.

Only the symmetrized coupling constants $\bar{B}_{kpq}=(B_{kpq}+B_{kqp})/2$ are relevant for
the hierarchy equations, Eq. \eqref{FOURIER_REL1} and \eqref{FIRST_PART}.
Several examples are calculated as outlined in the previous paragraphs,
\begin{eqnarray}
\nonumber
\bar{B}_{101}={1\over \pi}{\rm sin}\left({\alpha\over 2}\right)+
{1\over 2}\left[ {1\over 2}\left(1-{\alpha\over \pi}\right)-{\sin{\alpha}\over 2\pi}\right] \\
\nonumber
\bar{B}_{112}={1\over 6\pi}{\rm sin}\left({3\alpha\over 2}\right)+
{1\over 2}\left[ -{{\rm sin}(2\alpha)\over 8\pi}-{\sin{\alpha}\over 4\pi}\right] \\
\nonumber
\bar{B}_{123}={1\over 10\pi}{\rm sin}\left({5\alpha\over 2}\right)+
{1\over 2}\left[ -{{\rm sin}(3\alpha)\over 12\pi}-{{\rm sin}(2\alpha)\over 8\pi}\right] \\
\nonumber
\bar{B}_{211}={1\over 4\pi}(\alpha-
\sin{\alpha}) \\
\nonumber
\bar{B}_{202}={1\over 2\pi}\sin{\alpha}+
{1\over 2}\left[ -{{\rm sin}(2\alpha)\over 4\pi}+{1\over 2}\left(1-{\alpha\over \pi}\right) \right] \\
\nonumber
\bar{B}_{231}={1\over 8\pi}{\rm sin}\left(2\alpha\right)+
{1\over 2}\left[ -{{\rm sin}(\alpha)\over 4\pi}-{{\rm sin}(3\alpha)\over 12\pi}\right] \\
\nonumber
\bar{B}_{303}={1\over 3\pi}{\rm sin}\left({3\alpha\over 2}\right)+
{1\over 2}\left[ -{{\rm sin}(3\alpha)\over 6\pi}+{1\over 2}\left(1-{\alpha\over \pi}\right) \right] \\
\label{APPENDIX_BLAST}
\bar{B}_{312}={1\over 2\pi}{\rm sin}\left({\alpha\over 2}\right)+
{1\over 2}\left[ -{{\rm sin}(\alpha)\over 4\pi}-{{\rm sin}(2\alpha)\over 8\pi}\right] \:
\end{eqnarray}
The terms in the $[...]$ brackets are due to the asymmetric parts $B^{(1)}_{kpq}$ and vanish in the regular Vicsek limit of $\alpha=\pi$.
The coupling coefficients for the nematic theory of chapter \ref{sec:nematic} are
\begin{eqnarray}
\nonumber
& &\bar{B}_{224}={{\rm sin}(3\alpha)\over 12 \pi}
             -{{\rm sin}(2\alpha)\over 16 \pi}
             -{{\rm sin}(4\alpha)\over 32 \pi} \\
\nonumber
& &\bar{B}_{246}={{\rm sin}(5\alpha)\over 20 \pi}
             -{{\rm sin}(6\alpha)\over 48 \pi}
             -{{\rm sin}(4\alpha)\over 32 \pi} \\
\nonumber
& &\bar{B}_{404}={{\rm sin}(2\alpha)\over  4 \pi}
             -{{\rm sin}(4\alpha)\over 16 \pi}
            +{1\over 4}\left(1-{\alpha\over \pi}\right)\\
\nonumber
& &\bar{B}_{422}=-{{\rm sin}(2\alpha)\over  8 \pi}
              +{\alpha\over 4 \pi}\\
\nonumber
& &\bar{B}_{426}={{\rm sin}(4\alpha)\over 16 \pi}
             -{{\rm sin}(6\alpha)\over 48 \pi}
             -{{\rm sin}(2\alpha)\over 16 \pi} \\
\nonumber
& &\bar{B}_{624}={{\rm sin}( \alpha)\over  4 \pi}
             -{{\rm sin}(4\alpha)\over 32 \pi}
             -{{\rm sin}(2\alpha)\over 16 \pi} \\
\label{APPENDIX_NEMATIC}
& &\bar{B}_{606}={{\rm sin}(3\alpha)\over  6 \pi}
             -{{\rm sin}(6\alpha)\over 24 \pi}
            +{1\over 4}\left(1-{\alpha\over \pi}\right)\:
\end{eqnarray}

\section{Phase transitions in small systems}

Traditionally, phase transitions were only defined in equilibrium
and in the thermodynamic limit, $N\rightarrow \infty$, $V\rightarrow \infty$, $N/V=const.$
According to the oldest classification -- the Ehrenfest scheme  -- a macroscopic phase transition is indicated
by non-analytic behavior of the Gibbs free energy.

Since then, fueled by experiments and computer simulations,
the concept of phase transitions has been extended to far-from-equilibrium systems and small systems,
see for example
\cite{dunkel.j:2006,pleimling.m:2005,gross.d.h.e:2000,borrmann.p:2000,ihle.t:1994}.
This is especially relevant for systems where the thermodynamic limit
is not applicable or makes no sense experimentally,
such as in bird flocks, atomic nuclei, astrophysical objects or Bose-Einstein condensates
of atoms in a harmonic trap \cite{anderson.m.h:1995}.
Nevertheless,
even in systems with only $10^2-10^7$ particles,
it is sometimes possible to observe phenomena which are typical for phase transitions.
These phenomena can be but are not always precursors of phase transitions in the corresponding infinite system.
For example,
there are even equilibrium systems such as sodium clusters \cite{ellert.c:1997} where the nature of the
phase transition seems to change
with increasing particle number.
In spite of that, it has been shown that phase transitions can be classified at
fixed and finite particle number, avoiding the thermodynamic limit.
For example, Gross and Votyakov \cite{gross.d.h.e:2000} propose a classification of phase transitions in small systems
based on the topology of the Boltzmann entropy. Borrmann {\em et al.} \cite{borrmann.p:2000}
extract the type of phase transition
in small systems from the distribution of zeros of the partition function in the complex temperature plane.

Given that, at least in equilibrium, there exists such
a classification,
it seems appropriate to use the term ``phase transition'' for phase transition-like behavior
in the VM at
finite particle number.

The mean-field theory of Chapter ~\ref{sec:kinetic_theory} was derived in the thermodynamic limit.
The finding that its predictions for a tricritical point agree {\em quantitatively} with
agent-based simulations in moderately small boxes further justifies the idea to speak of
phase transitions and tricritical points in small systems.

\vfil

\bibliography{references}

\begin{thebibliography}{60}
\bibitem[{FOO({\natexlab{a}})}]{FOOT2.ihle:2014}
\bibinfo{note}{$\sin{\theta}$ terms would be needed if either the angular noise
  would not be symmetric with respect to positive and negative values of the
  angle or if the local mean flow direction would differ from the
  $x$-direction}.
\bibitem[{FOO({\natexlab{b}})}]{FOOT3.ihle:2014}
\bibinfo{note}{The noise $\eta$ in our system loosely corresponds to
  temperature in a condensed matter system because there is a transition from a
  disordered to an ordered state if the noise is reduced}.
\end{thebibliography}

\end{document}